\documentclass[%
 aip,
 amsmath,amssymb,
 reprint,%
]{revtex4-2}

\usepackage{graphicx}
\usepackage{dcolumn}
\usepackage{bm}

\usepackage[utf8]{inputenc}
\usepackage[T1]{fontenc}
\usepackage{mathptmx}
\usepackage{etoolbox}
\DeclareMathAlphabet{\mathdutchcal}{U}{dutchcal}{m}{n}
\SetMathAlphabet{\mathdutchcal}{bold}{U}{dutchcal}{b}{n}
\DeclareMathAlphabet{\mathdutchbcal}{U}{dutchcal}{b}{n}
\usepackage{mathtools}
\usepackage{mathrsfs}
\usepackage{upgreek}
\usepackage{stmaryrd}
\usepackage{xcolor}

\usepackage{scalerel}

\makeatletter
\def\@email#1#2{%
 \endgroup
 \patchcmd{\titleblock@produce}
  {\frontmatter@RRAPformat}
  {\frontmatter@RRAPformat{\produce@RRAP{*#1\href{mailto:#2}{#2}}}\frontmatter@RRAPformat}
  {}{}
}%
\makeatother
\begin{document}

\preprint{AIP/123-QED}

\title[Microscopic Derivation of Stochastic Thin-Film Equation]{Microscopic Derivation of Stochastic Thin-Film Equation and
Extensions to Two-Layer and Evaporating Films}
\author{Toby Kay}
 \email{t.kay@imperial.ac.uk, s.kalliadasis@imperial.ac.uk}
\author{Serafim Kalliadasis}%
\affiliation{
Complex Multiscale Systems Group, Department of Chemical
Engineering, Imperial College, London,
SW7 2AZ, United Kingdom
}%

\date{\today}

\begin{abstract}
While classical thin-film hydrodynamics models widely govern macroscopic
fluid layers, they fail to capture the molecular-level thermal fluctuations
that dictate nanoscale interfacial dynamics. Standard stochastic extensions
rely on the long-wave approximation of the continuum fluctuating
Navier–Stokes equations; however, this top-down paradigm lacks a formal
connection to microscopic particle trajectories and cannot readily
accommodate complex, non-equilibrium boundary mechanisms. We resolve this
limitation by deriving the stochastic thin-film equation directly from the
exact underlying \(N\)-particle Hamiltonian dynamics. This is achieved via
a systematic, bottom-up coarse-graining framework based on a generalized
fluctuating dynamic density-functional theory. Crucially, this
first-principles formulation offers unparalleled flexibility, enabling the
seamless integration of multi-layered interfaces and non-conserved dynamics
such as evaporation. For these complex configurations, our framework
rigorously yields entirely novel stochastic thin-film equations featuring
emergent noise terms that encapsulate microscopic fluctuations missing from
classical top-down models.
\end{abstract}

\maketitle

\section{Introduction}

Liquid films are ubiquitous across the natural world and span vastly
separated scales, ranging from physiological microenvironments, such as the
alveolar lining of the lungs, to massive geophysical processes typified by
lava flows. Interfacial layers with characteristic thicknesses spanning from
a few nanometers up to a micron are of profound technological and fundamental
interest. They govern a diverse array of biological and engineering systems,
including thin corneal films, falling liquid layers, subaquatic gravity
currents, industrial protective coatings, falling-film reactors,
microelectronic cooling arrays, and the rapidly expanding domain of micro-
and
nanofluidics~\cite{oron1997long,Kalliadasis2012falling,hermans2015lung,piegari2018optical}.

The mathematical modeling of thin-film hydrodynamics conventionally follows
two paradigms. The first utilises a top-down reduction of the classical
macroscopic Navier--Stokes equations and their associated boundary conditions
via a long-wave expansion based on the lubrication
approximation~\cite{oron1997long,miksis1998dynamics,Kalliadasis2012falling},
tracing back to the seminal formulation by Reynolds~\cite{reynolds1885theory}
(see Refs.~\onlinecite{oron1997long,Kalliadasis2012falling} for reviews). The
second paradigm relies on a phenomenological variational gradient-descent
framework, wherein thin-film evolution is driven by the minimisation of an
interfacial free-energy
functional~\cite{thiele2011note,thiele2013gradient,thiele2014patterned,thiele2018recent}.
Crucially, because both treatments operate strictly at the macroscopic limit,
they lack a formal connection to microscopic molecular trajectories. To
bridge this gap, te Vrugt {\it{et al.}}~\onlinecite{te2024microscopic}
recently derived a deterministic thin-film equation from first-principles
Hamiltonian dynamics by utilising the Mori--Zwanzig projection operator
technique~\cite{zwanzig1961memory,Mori1965,te2020projection}, yielding a
generalised dynamic density-functional theory (DDFT)
approach~\cite{te2020classical}.

Nevertheless, these existing frameworks remain fundamentally deterministic,
meaning microscopic fluctuations are averaged out. At the nanoscale, this
omission is unphysical; thermal fluctuations exert a dominant influence on
interfacial dynamics. Molecular fluctuations manifest as spontaneous
stochastic forces that alter the local film height, accelerating topological
transitions, triggering film breakup, and dramatically modulating dewetting
pathways~\cite{grun2006thin,duran2019instability}. While a stochastic
thin-film (STF) equation can be derived via a long-wave expansion of the
phenomenological Landau--Lifshitz fluctuating Navier--Stokes
equations~\cite{duran2019instability,landau1987fluid} or via a variational
gradient descent coupled with the fluctuation--dissipation
relation~\cite{schaefer2016structuring,de2024structuring}, these approaches
still lack a bottom-up micro-to-macro linkage because their continuum
starting equations are themselves coarse-grained.

We overcome this restriction by deriving the STF equation directly from exact
microscopic dynamics. While our projection operator approach parallels the
mathematical structure of Ref.~\onlinecite{te2024microscopic}, our
formulation operates strictly at the mesoscopic—and not macroscopic—level,
thereby rigorously preserving thermal fluctuations. To achieve this, we
deploy a generalised fluctuating dynamic density-functional theory (FDDFT)
framework recently established by us in Ref.~\onlinecite{Kay2025generalised},
which generalizes our earlier classical fluid
formulation~\cite{duran2017general} to arbitrary slow observables. This
systematic, first-principles strategy maps the stochastic evolution of
mesoscopic variables onto a central structural functional (entropy or free
energy). As demonstrated below, this bottom-up approach natively recovers the
STF equation while offering an unmatched mathematical flexibility to
incorporate increasing levels of physical complexity seamlessly.

\section{Review of FDDFT}\label{sec:fddft}

To provide context, we review the generalised FDDFT framework developed by us
in Ref.~\onlinecite{Kay2025generalised}, which serves as our foundational
statistical-mechanics vehicle for deriving the stochastic thin-film equation
and its extensions.

DDFT extends classical DFT to non-equilibrium configurations under
time-dependent dynamics. In its conventional formulation, DDFT dictates that
the spatial density field of a classical fluid evolves down the functional
gradient of an interfacial free energy. It has been deployed across a host of
applications—including colloidal suspensions, molecular fluids, polymers,
thin films, and glassy systems—to explore phenomena ranging from phase
separation and pattern formation to nucleation dynamics and chemical
reactions (reviews by Lutsko~\onlinecite{Lutsko2010} and te Vrugt {\emph{et
al.}}~\onlinecite{te2020classical}). Subsequent modifications have
incorporated hydrodynamic interactions~\cite{rex2008dynamical}, inertial
coupling~\cite{goddard2012general,goddard2013unification}, particle
non-sphericity~\cite{duran2016dynamical}, multi-component
mixtures~\cite{goddard2013multi}, and thermal
gradients~\cite{wittkowski2012extended}. Consequently, generalised DDFT
formulations derived via projection operator formalisms focus on generic slow
observables and their respective mesoscopic evolution
paths~\cite{grabert2006projection,espanol2009derivation,wittkowski2012extended,duran2017general,te2020projection,Kay2025generalised}.

However, conventional DDFT remains strictly macroscopic, tracking
ensemble-averaged microscopic fields wherein spontaneous fluctuations are
systematically neglected. It is thus inherently ill-suited for mesoscopic
regimes, such as nanoscale liquid layers, where thermal noise dictates the
interfacial topology. In this context, the mesoscale defines a separation of
scales: it is sufficiently coarse to avoid tracking discrete molecular
trajectories, yet refined enough to preserve emergent thermal fluctuations.
Within this intermediate regime, fluctuating dynamic density-functional
theory becomes applicable.

Consider a generic $N$-particle, $n$-species Hamiltonian system. The
instantaneous microstate is defined by its phase-space coordinates
$\bm{\Gamma}_t \equiv
\{(\mathbf{q}^\alpha_{i}(t),\mathbf{p}^\alpha_{i}(t))\vert
i=1,\dots,N_\alpha,\,\alpha=1,\dots,n\}$, where $\mathbf{q}^\alpha_{i}$ and
$\mathbf{p}^\alpha_{i}$ denote the position and momentum of the $i$-th
particle of species $\alpha$, respectively, and $N_\alpha$ is the particle
count of species $\alpha$ such that $\sum_{\alpha=1}^n N_\alpha=N$. We write
the Hamiltonian in the general form $\hat{H}=\sum_{\alpha=1}^n
\sum_{i=1}^{N_\alpha}|\mathbf{p}^\alpha_{i}|^2/(2m_\alpha)+V(\{\mathbf{q}^\alpha\})$,
where hats denote microstate-dependent quantities.

For a designated fluid configuration, one identifies a set of resolved
microscopic observable fields $\hat{\mathbf{Z}}_i(\mathbf{r})$ parameterized
by the 3D position vector $\mathbf{r}$. By Liouville's theorem, the time
evolution of these fields is governed by the Liouvillian operator,
$\dot{\hat{\mathbf{Z}}_i} = i\mathcal{L}\hat{\mathbf{Z}}_i =
\{\hat{\mathbf{Z}}_i,\hat{H}\}$, defined via the standard Poisson bracket.
Utilising the Zwanzig projection operator technique~\cite{zwanzig1961memory}
under a clean separation of timescales between the slow resolved variables
and the fast unresolved degrees of freedom, we obtain the generalized FDDFT
framework~\cite{Kay2025generalised}:
\begin{align}\label{eq:fddft_entropy}
    \partial_t \mathbf{z}_i(\mathbf{r},t) &= \left\langle i \mathcal{L} \hat{\mathbf{Z}}_i (\mathbf{r})\right\rangle_{\mathbf{z}} + \frac{1}{k_B}\sum_j \int \mathrm d\mathbf{r}' \mathbf{M}_{ij}[\mathbf{z};\mathbf{r},\mathbf{r}') \cdot \frac{\delta S[\mathbf{z}]}{\delta \mathbf{z}_j(\mathbf{r}',t)} \nonumber \\
    &+ \sum_j \int \mathrm d\mathbf{r}' \frac{\delta \mathbf{M}_{ij}[\mathbf{z};\mathbf{r},\mathbf{r}') }{\delta \mathbf{z}_j(\mathbf{r}',t)} + \sum_j \int \mathrm d\mathbf{r}' \mathbf{B}_{ij}[\mathbf{z};\mathbf{r},\mathbf{r}') \cdot \bm{\xi}_j(\mathbf{r}',t).
\end{align}
Equation~\eqref{eq:fddft_entropy} governs the mesoscopic observable defined
as the constrained ensemble average $\mathbf{z}_i(\mathbf{r},t)=\langle
\hat{\mathbf{Z}}_i(\mathbf{r})\rangle_{\mathbf{z}(\mathbf{r},t)}$. This
generalised microcanonical average of a generic phase-space variable
$\hat{X}$ is formulated as~\cite{zwanzig1961memory}:
\begin{equation}\label{eq:micro_ensemble_average}
    \langle \hat{X} \rangle_\mathbf{z}=\frac{c_N}{\Omega[\mathbf{z}]}\int \mathrm d\bm{\Gamma} \delta[\hat{\mathbf{Z}}-\mathbf{z}] \hat{X}(\bm{\Gamma}),
\end{equation}
with the density of states normalized via:
\begin{equation}\label{eq:omega}
    \Omega[\mathbf{z}]=c_N \int \mathrm d\bm{\Gamma} \delta[\hat{\mathbf{Z}}-\mathbf{z}].
\end{equation}
Here, the functional Dirac delta is defined as $\delta[\hat{\mathbf{Z}}-\mathbf{z}]=\prod_{i}^{M}\delta[\hat{\mathbf{Z}}_i-\mathbf{z}_i]$,
while $c_N=(\prod_{\alpha=1}^n N_\alpha ! h^{3N})^{-1}$ accounts for particle indistinguishability within each species, where $h$ is Planck's constant.
The right-hand side of Eq.~\eqref{eq:fddft_entropy} systematically partitions the mesoscopic dynamics into reversible (first term),
dissipative (second and third terms), and fluctuating (final term) contributions.

The structural entropy functional $S[\mathbf{z}]$ is related to the density of states by:
\begin{equation}
  S[\mathbf{z}] =  k_B \ln \left(\Omega[\mathbf{z}]\Delta \mathbf{z}\right),
\end{equation}
where $\Delta \mathbf{z}$ represents the characteristic volume width of the constrained macrostate manifold.

The operator $\mathbf{M}[\mathbf{z};\mathbf{r},\mathbf{r}')$ in
Eq.~\eqref{eq:fddft_entropy} denotes the dissipative tensor, defined in a
generalised Green--Kubo form as:
\begin{equation}\label{eq:Green_kubo_full}
    \mathbf{M}_{ij}[\mathbf{z};\mathbf{r},\mathbf{r}')=\int_0^\infty \mathrm ds \left\langle \mathcal{Q}i\mathcal{L}\hat{\mathbf{Z}}_j(\mathbf{r'}) \otimes \mathcal{Q}e^{i\mathcal{L}\mathcal{Q}s}i\mathcal{L}\hat{\mathbf{Z}}_i(\mathbf{r}) \right\rangle_\mathbf{z},
\end{equation}
where $\mathcal{Q}=\mathcal{I}-\mathcal{P}$ is the orthogonal complement of
the Zwanzig projection operator $\mathcal{P}\hat{X}=\langle
\hat{X}\rangle_{\hat{\mathbf{Z}}}$ and $\mathcal{I}$ is the identity. The
dissipative tensor is tied to the stochastic noise amplitude
$\mathbf{B}[\mathbf{z};\mathbf{r},\mathbf{r}')$ through the
fluctuation--dissipation relation:
\begin{equation}\label{eq:fluctuation_dissipation}
    \int \mathrm d\mathbf{r}'' \mathbf{B}[\mathbf{z};\mathbf{r},\mathbf{r}'')\mathbf{B}^\top[\mathbf{z};\mathbf{r}',\mathbf{r}'')=2 \mathbf{M}^S[\mathbf{z};\mathbf{r},\mathbf{r}'),
\end{equation}
where superscript $S$ signifies the symmetric part. The spatio-temporal white
noise $\bm{\xi}_i(\mathbf{r},t)$ is mean-zero with the correlation structure:
\begin{equation}
      \mathbb{E}\left[\bm{\xi}_i(\mathbf{r},t)\otimes \bm{\xi}_j(\mathbf{r}',t')\right]=\delta_{ij}\delta(\mathbf{r}-\mathbf{r}')\delta(t-t')\mathbf{I}.
\end{equation}

By leveraging the scale separation inherent to this mesoscopic description,
we can invoke an ultraviolet wavenumber cut-off ubiquitous in fluctuating
hydrodynamics~\cite{forster1976long,forster1977large,glorioso2018lectures,yokota2026unifying,nakano2025looking}.
This spectral truncation excludes high-wavenumber microscopic modes that
dictate the long-time hydrodynamic tails of Eq.~\eqref{eq:Green_kubo_full}.
Consequently, the upper integration limit can be truncated at a mesoscopic
timescale $\tau$ that is sufficiently long to capture short-time molecular
correlations but short compared to the long-time relaxation of modes beyond
the cut-off. Over this interval, the resolved variables evolve slowly,
justifying the operator approximation $e^{i\mathcal{L}s}\approx
e^{i\mathcal{L}\mathcal{Q}s}$. This reduces the dissipative tensor to the
localised form:
\begin{equation}
    \mathbf{M}_{ij}[\mathbf{z};\mathbf{r},\mathbf{r}')=\int_0^\tau \mathrm ds \left\langle \mathcal{Q}\dot{\hat{\mathbf{Z}}}_j(\mathbf{r'},0) \otimes  \mathcal{Q}\dot{\hat{\mathbf{Z}}}_i(\mathbf{r},s) \right\rangle_\mathbf{z},
\end{equation}
where the instantaneous flux satisfies
$\dot{\hat{\mathbf{Z}}}_i(\mathbf{r},s)=e^{i\mathcal{L}s}\dot{\hat{\mathbf{Z}}}_i(\mathbf{r},0)$.

In an isothermal system, the Hamiltonian $\hat{H}$ is implicitly included
among the observables, though its invariance ($i\mathcal{L}\hat{H}=0$)
ensures it modulates only the ensemble weights rather than entering the flux
dynamics directly. When focusing strictly on momentum-independent fields
(implying a zero bulk flow condition), the internal energy reduces to
$U=\langle \hat{H} \rangle_\mathbf{z}$. Assuming local thermodynamic
equilibrium holds under sufficiently slow macrostate evolution, we
approximate the microcanonical average via the local equilibrium
ensemble~\cite{zubarev1973nonequilibrium}:
\begin{equation}\label{eq:gen_canonical_Hamiltonian}
  \langle \hat{X} \rangle_\text{leq}^t=\frac{c_N}{\mathcal{Z}[\bm{\lambda}]}\int \mathrm d\bm{\Gamma} \hat{X} \exp\left(-\beta \hat{H}-k_B^{-1}\sum_i \int \mathrm d\mathbf{r} \bm{\lambda}_i(\mathbf{r},t) \cdot \hat{\mathbf{Z}}_i \right),
\end{equation}
where $\beta=(k_BT)^{-1}$ and the Lagrange multipliers
$\bm{\lambda}_i(\mathbf{r},t)=\delta S/\delta \mathbf{z}_i$ guarantee
consistency, i.e., $\langle
\hat{\mathbf{Z}}_i(\mathbf{r})\rangle_\text{leq}^t=\mathbf{z}_i(\mathbf{r},t)$.
Utilising the thermodynamic relation $F=U-TS$, the functional derivative of
the entropy reads:
\begin{equation}
  \frac{\delta S[\mathbf{z}]}{\delta \mathbf{z}_i(\mathbf{r},t)} = -\frac{1}{T}\frac{\delta F[\mathbf{z}]}{\delta \mathbf{z}_i(\mathbf{r},t)},
\end{equation}
where $F[\mathbf{z}]$ is the free-energy functional.

Furthermore, for momentum-independent observables, the reversible transport
fluxes vanish identically ($\left\langle i \mathcal{L} \hat{\mathbf{Z}}_i
(\mathbf{r})\right\rangle_\text{leq}^t=0$) because the operator $i
\mathcal{L} \hat{\mathbf{Z}}_i (\mathbf{r})$ is strictly linear in momentum,
whereas the local canonical ensemble~\eqref{eq:gen_canonical_Hamiltonian}
remains symmetric (Gaussian) in momentum space. Consequently, the mesoscopic
isothermal FDDFT for momentum-independent variables simplifies to the pure
gradient-descent form:
\begin{align}\label{eq:fddft_free_energy}
    \partial_t \mathbf{z}_i(\mathbf{r},t)&=- \beta \sum_j \int \mathrm d\mathbf{r}' \mathbf{M}_{ij}(\mathbf{r},\mathbf{r}',t) \cdot \frac{\delta F[\mathbf{z}]}{\delta \mathbf{z}_j(\mathbf{r}',t)} \nonumber \\
    &+ \sum_j \int \mathrm d\mathbf{r}' \frac{\delta \mathbf{M}_{ij}(\mathbf{r},\mathbf{r}',t) }{\delta \mathbf{z}_j(\mathbf{r}',t)} + \sum_j \int \mathrm d\mathbf{r}' \mathbf{B}_{ij}(\mathbf{r},\mathbf{r}',t) \cdot \bm{\xi}_j(\mathbf{r}',t),
\end{align}
where the time-dependent localised dissipative tensor is explicitly closed
by:
\begin{equation}\label{eq:dissipative_tensor}
  \mathbf{M}_{ij}(\mathbf{r},\mathbf{r}',t)=\int_0^\tau \mathrm ds \left\langle \dot{\hat{\mathbf{Z}}}_j(\mathbf{r'},0) \otimes  \dot{\hat{\mathbf{Z}}}_i(\mathbf{r},s) \right\rangle_\text{leq}^t.
\end{equation}
Equation~\eqref{eq:fddft_free_energy} serves as our definitive,
structure-preserving starting point for deriving the stochastic thin-film
equations and their non-equilibrium extensions.

\section{Derivation of the STF equation}\label{sec:stf}

\begin{figure}[htbp]
  \centering
  \includegraphics[width=0.45\textwidth]{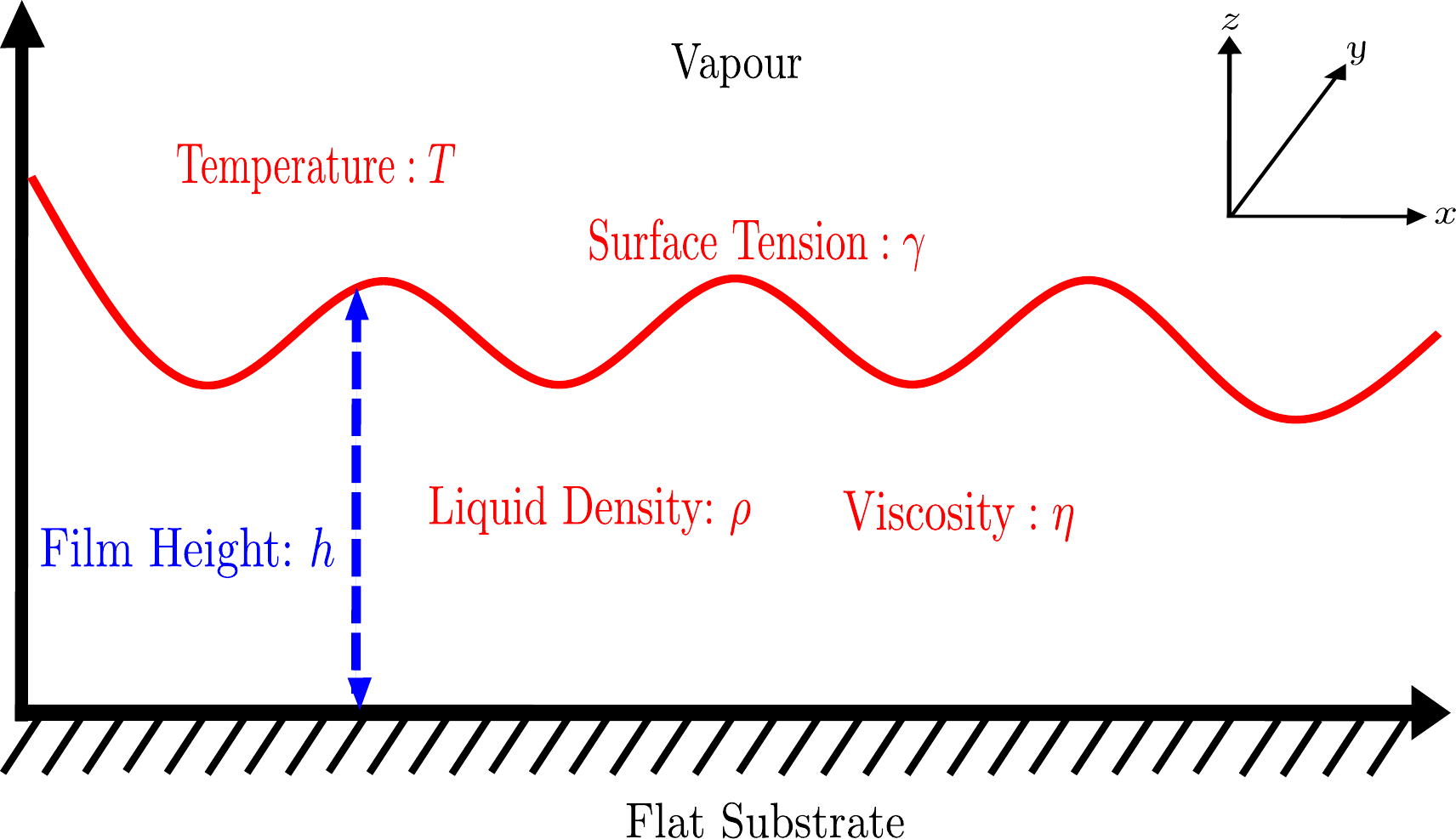}
  \caption{Sketch of the profile geometry of a thin liquid film of height $h$. The streamwise and cross-stream coordinates are given by $x$ and $y$, respectively, with the substrate coinciding with the $(x,y)$ plane, and $z$ denoting the outward-pointing coordinate normal to the substrate.}
  \label{fig:film}
\end{figure}

We consider a thin film of an incompressible fluid bounded by a flat solid
substrate parallel to the $(x,y)$ plane at $z=0$ within a Cartesian
coordinate system $(x,y,z)$. Figure~\ref{fig:film} sketches the physical
configuration, where $x$ and $y$ denote the streamwise and spanwise
coordinate directions, respectively, and $z$ is the outward-pointing
coordinate normal to the solid boundary. The instantaneous local film height
$z=h(\mathbf{u},t)$ defines the slow mesoscopic observable of interest,
parameterized by the 2D lateral position vector $\mathbf{u}=(x,\, y)^\top$,
while the full 3D position vector is $\mathbf{r}=(x,\, y,\, z)^\top$. To
formulate the stochastic transport dynamics for $h(\mathbf{u},t)$ via the
generalised FDDFT~\eqref{eq:fddft_free_energy}, we introduce a microscopic
height field $\hat{h}(\mathbf{u})$ whose constrained ensemble average matches
the continuous profile, $\langle
\hat{h}(\mathbf{u})\rangle_\text{leq}^t=h(\mathbf{u},t)$. Following
Refs.~\onlinecite{Antonio2012,Andreas2014}, the microscopic film height is
rigorously related to the microscopic fluid density field
$\hat{\rho}(\mathbf{r})$ via:
\begin{equation}
    \hat{h}(\mathbf{u})=\frac{1}{\rho_l-\rho_g}\int_0^\infty [\hat{\rho}(\mathbf{r})-\rho_g]\,\mathrm dz,
\end{equation}
where $\rho_l$ and $\rho_g$ denote the liquid and surrounding gas phase
densities, respectively. Under the sharp phase-density separation limit
$\rho_l\gg \rho_g$, this definition simplifies asymptotically
to~\cite{te2024microscopic}:
\begin{equation}\label{eq:microscopic_height}
    \hat{h}(\mathbf{u}) \simeq \frac{1}{\rho_l}\int_0^\infty \hat{\rho}(\mathbf{r})\,\mathrm dz,
\end{equation}
where the microscopic density operator is defined over $N$ discrete particles
as $\hat{\rho}(\mathbf{r})=\sum_{i=1}^N \delta(\mathbf{r}-\mathbf{q}_i)$.

Evaluating the free-energy-driven FDDFT~\eqref{eq:fddft_free_energy} for the
resolved height field requires closing the generalized dissipative
tensor~\eqref{eq:dissipative_tensor} for $\hat{\mathbf{Z}}=\hat{h}$ by
extending the micro-scaling procedure of Ref.~\onlinecite{te2024microscopic}.
The exact microstate temporal flux of the film height satisfies the
continuous conservation law:
\begin{equation}
    \dot{\hat{h}}(\mathbf{u})=i\mathcal{L} \hat{h}(\mathbf{u})=-\bm{\nabla}_{\mathbf{u}}\cdot \hat{\mathbf{j}}^{\parallel}(\mathbf{u}),
\end{equation}
where $\hat{\mathbf{j}}^{\parallel}(\mathbf{u})$ is the 2D microscopic
current density parallel to the substrate boundary:
\begin{equation}
    \hat{\mathbf{j}}^{\parallel}(\mathbf{u})=\frac{1}{\rho_l}\int_0^\infty \mathrm dz \sum_{i=1}^N (\upsilon_i^x,\,\upsilon_i^y)^\top \delta(\mathbf{r}-\mathbf{q}_i),
\end{equation}
with $\bm{\upsilon}_i=\mathbf{p}_i/m$ representing individual particle
velocities. Projecting these spatial fluxes maps the height dissipative
tensor into the dual-divergence form:
\begin{equation}
    M(\mathbf{u},\mathbf{u}',t)=\bm{\nabla}_{\mathbf{u}}\cdot \bm{\nabla}_{\mathbf{u}'}\cdot \int_0^\tau \mathrm ds \left\langle \hat{\mathbf{j}}^{\parallel}(\mathbf{u}') \otimes \hat{\mathbf{j}}^{\parallel}(\mathbf{u},s)\right\rangle_\text{leq}^t.
\end{equation}
We define the underlying momentum Onsager matrix tensor as the current-flux
autocorrelation integral~\cite{te2024microscopic}:
\begin{equation}
  \mathbf{D}(\mathbf{u},\mathbf{u}',t) = \int_0^\tau \mathrm ds \left\langle \hat{\mathbf{j}}^{\parallel}(\mathbf{u}') \otimes \hat{\mathbf{j}}^{\parallel}(\mathbf{u},s)\right\rangle_\text{leq}^t,
\end{equation}
such that $M(\mathbf{u},\mathbf{u}',t)=\bm{\nabla}_{\mathbf{u}}\cdot
\bm{\nabla}_{\mathbf{u}'}\cdot \mathbf{D}(\mathbf{u},\mathbf{u}',t)$.
Introducing the vertically resolved parallel current flux
$\hat{\mathbf{j}}^{\parallel}(\mathbf{u},z)=\sum_i
(\upsilon_i^x,\,\upsilon_i^y)^\top \delta(\mathbf{r}-\mathbf{q}_i)/\rho_l$
reveals that the total current is an integrated marginal projection,
$\hat{\mathbf{j}}^{\parallel}(\mathbf{u})=\int \mathrm dz
\hat{\mathbf{j}}^{\parallel}(\mathbf{u},z)$, yielding:
\begin{equation}\label{eq:marginal_onsager}
  \mathbf{D}(\mathbf{u},\mathbf{u}',t)=\int_0^\infty \mathrm dz \int_0^\infty \mathrm dz' \mathbf{D}(\mathbf{u},z,\mathbf{u}',z',t).
\end{equation}
Invoking lateral isotropic rotational invariance within the substrate plane
allows us to decompose the current tensor as:
\begin{equation}
    \langle \hat{\mathbf{j}}^{\parallel}(\mathbf{u}',z') \otimes \hat{\mathbf{j}}^{\parallel}(\mathbf{u},z,s)\rangle_\text{leq}^t = \frac{1}{2}\langle \hat{\mathbf{j}}^{\parallel}(\mathbf{u}',z') \cdot \hat{\mathbf{j}}^{\parallel}(\mathbf{u},z,s)\rangle_\text{leq}^t \mathbf{I},
\end{equation}
where $\mathbf{I}$ is the $2\times 2$ identity matrix, which simplifies the
kernels to $\mathbf{D}(\mathbf{u},z,\mathbf{u}',z',t)=\frac{1}{2}\int_0^\tau
\mathrm ds \langle \hat{\mathbf{j}}^{\parallel}(\mathbf{u}',z') \cdot
\hat{\mathbf{j}}^{\parallel}(\mathbf{u},z,s)\rangle_\text{leq}^t \mathbf{I}$.
Integrating this kernel over the total interface surface domain isolates the
transverse current correlation function $C(z,z',t,s)$:
\begin{align}
  \int \mathrm d\mathbf{u} \int \mathrm d\mathbf{u}' \mathbf{D}(\mathbf{u},z,\mathbf{u}',z',t) &= \frac{1}{2} \int_0^\tau \mathrm ds \langle \hat{\mathbf{j}}^{\parallel}(z') \cdot \hat{\mathbf{j}}^{\parallel}(z,s)\rangle_\text{leq}^t \mathbf{I} \nonumber \\
  &= \int_0^\tau \mathrm ds C(z,z',t,s) \mathbf{I},
\end{align}
where $\hat{\mathbf{j}}^{\parallel}(z)=\int
\hat{\mathbf{j}}^{\parallel}(\mathbf{u},z)\,\mathrm d\mathbf{u}$. Finally,
enforcing a strict spatial locality approximation within the lateral planes
dictates that parallel velocity correlations vanish identically for any
distinct positions $\mathbf{u}\neq\mathbf{u}'$, yielding the delta-localised
identification~\cite{te2024microscopic}:
\begin{align}\label{eq:onsager_transverse}
  \mathbf{D}(\mathbf{u},z,\mathbf{u}',z',t)&=\frac{1}{2A} \delta(\mathbf{u}-\mathbf{u}')\int_0^\tau \mathrm ds \langle \hat{\mathbf{j}}^{\parallel}(z') \cdot \hat{\mathbf{j}}^{\parallel}(z,s)\rangle_\text{leq}^t \mathbf{I} \nonumber \\
  &=A^{-1}\delta(\mathbf{u}-\mathbf{u}')\int_0^\tau \mathrm ds C(z,z',t,s) \mathbf{I},
\end{align}
where $A=\int \mathrm d\mathbf{u}$ defines the macroscopic substrate surface
area. Computing the algebraic closure for the thin-film Onsager matrix is
thus fundamentally reduced to evaluating the transverse current correlation
function, $C(z,z',t,s)$.

Via Onsager's regression hypothesis of
fluctuations~\cite{onsager1931reciprocal}, the transverse current correlation
function under near-equilibrium conditions satisfies the continuous
hydrodynamic diffusion
equation~\cite{palmer1994transverse,bocquet1994hydrodynamic,hess2002determining,hansen2013theory,jakse2013liquid,ropo2016collective,te2024microscopic}:
\begin{equation}\label{eq:diffusion}
    \partial_s C(z,z',t,s) = \nu \partial_z^2 C(z,z',t,s),
\end{equation}
where $\nu=\eta/(m\rho_l)$ is the kinematic viscosity and $\eta$ is the
dynamic viscosity. Utilising the canonical
ensemble~\eqref{eq:gen_canonical_Hamiltonian}, we establish the initial
condition for the correlation evolution via the split relation:
\begin{align}
    C(z,z',t,0) &=\frac{1}{2 \rho_l^2}\left\langle \sum_{ij} \bm{\upsilon}_i \cdot \bm{\upsilon}_j \delta(z-q_i^z) \delta(z'-q_j^z) \right\rangle_\text{leq}^t \nonumber \\
    &=\frac{1}{2 \rho_l^2}\left\langle \sum_{i} \bm{\upsilon}_i \cdot \bm{\upsilon}_i \delta(z-q_i^z) \delta(z'-z) \right\rangle_\text{leq}^t \nonumber \\
    &=\frac{\langle \hat{\rho}(z)\rangle_\text{leq}^t}{\beta m \rho_l^2}  \delta(z-z') \mathbb{I}_{0<z<h},
\end{align}
where $\mathbb{I}_{0<z<h}$ is the indicator function over the film thickness.
Setting $\langle \hat{\rho}(z)\rangle_\text{leq}^t=\rho_l A$, we exploit the
delta localisation in $z'$ to define the marginal quantity:
\begin{equation}\label{eq:marginal_transverse_correlation}
  \mathcal{C}(z,t,s)=\int_0^\infty \mathrm dz' \frac{m \rho_l \beta}{A} C(z,z',t,s),
\end{equation}
which satisfies the simplified initial state
$\mathcal{C}(z,t,0)=\mathbb{I}_{0<z<h}$. At the solid substrate boundary
($z=0$), we enforce a hydrodynamic no-slip condition forcing the parallel
current correlation to vanish ($C(0,z',t,s)=0$),
yielding~\cite{bocquet1994hydrodynamic}:
\begin{equation}\label{eq:z0_bc}
  \mathcal{C}(0,t,s)=0.
\end{equation}
At the unconstrained liquid-vapor free surface ($z=h$), the stagnant vapor
phase exerts negligible shear, prescribing a vanishing tangential stress
boundary condition ($\partial_z C(h,z',t,s)=0$), hence:
\begin{equation}\label{eq:zh_bc}
  \partial_z \mathcal{C}(h,t,s)=0.
\end{equation}

To solve the bounded diffusion problem~\eqref{eq:diffusion}, we implement the
Laplace transform technique, defined for a generic time-dependent field
$f(s)$ as $\mathcal{L}\{f(s)\}=\widetilde{f}(\epsilon)=\int_0^\infty \mathrm
ds e^{-\epsilon s} f(s)$. Transforming Eq.~\eqref{eq:diffusion} subject to
the initial state~\eqref{eq:initial_condition} provides the ordinary
differential equation:
\begin{equation}
  \epsilon \widetilde{\mathcal{C}}(z,t,\epsilon)-1=\nu \partial^2_z \widetilde{\mathcal{C}}(z,t,\epsilon),
\end{equation}
which is analytically closed under boundary conditions~\eqref{eq:z0_bc}
and~\eqref{eq:zh_bc} as:
\begin{equation}
  \widetilde{\mathcal{C}}(z,t,\epsilon) = \frac{-e^{\frac{\sqrt{\epsilon } (2 h-z)}{\sqrt{\nu}}}+e^{\frac{2 h \sqrt{\epsilon }}{\sqrt{\nu}}}-e^{\frac{z \sqrt{\epsilon }}{\sqrt{\nu}}}+1}{\epsilon  \left(e^{\frac{2 h \sqrt{\epsilon }}{\sqrt{\nu}}}+1\right)}.
\end{equation}
Combining
relations~\eqref{eq:marginal_onsager},~\eqref{eq:onsager_transverse},
and~\eqref{eq:marginal_transverse_correlation} yields the integrated parallel
current description:
\begin{equation}\label{eq:transverse_corr}
    \mathbf{D}(\mathbf{u},\mathbf{u}',t)=\frac{k_B T \nu}{\eta} \delta(\mathbf{u}-\mathbf{u}')\int_0^{h}\mathrm dz \int_0^\tau \mathrm ds \mathcal{C}(z,t,s) \mathbf{I}.
\end{equation}
Because the mesoscopic cut-off timescale $\tau$ vastly exceeds the
characteristic molecular correlation length of $\mathcal{C}(z,t,s)$, we
evaluate the asymptotic limit $\tau\to \infty$. Utilising the Laplace
transform final-value theorem~\cite{roberts1966table}, $\lim_{t\to \infty}
f(t)=\lim_{\epsilon \to 0} \epsilon\widetilde{f}(\epsilon)$, alongside the
integral transform identity $\mathcal{L}\left\{ \int_0^t \mathrm ds f(s)
\right\}=\widetilde{f}(\epsilon)/\epsilon$, we isolate the long-time
correlation metric:
\begin{equation}\label{eq:laplace_limit}
  \lim_{\tau \to \infty} \int_0^\tau \mathrm ds \mathcal{C}(z,t,s)=\lim_{\epsilon \to 0} \widetilde{\mathcal{C}}(z,t,\epsilon)=\frac{z (2 h-z)}{2 \nu}.
\end{equation}
Substituting Eq.~\eqref{eq:laplace_limit} into Eq.~\eqref{eq:transverse_corr}
explicitly recovers the classic mobility-scaled Onsager matrix:
\begin{equation}\label{eq:stf_onsager_matrix}
  \mathbf{D}(\mathbf{u},\mathbf{u}',t)=k_B T \frac{h^3(\mathbf{u},t)}{3 \eta} \delta(\mathbf{u}-\mathbf{u}') \mathbf{I},
\end{equation}
which directly maps the macroscopic height dissipative tensor into the
double-divergence format:
\begin{equation}\label{eq:STF_dissipative_tensor}
  M(\mathbf{u},\mathbf{u}',t)=k_B T \bm{\nabla}_{\mathbf{u}}\cdot\frac{h^3(\mathbf{u},t)}{3 \eta} \bm{\nabla}_{\mathbf{u}'}\delta(\mathbf{u}-\mathbf{u}').
\end{equation}
Owing to the spatial divergence structure of
Eq.~\eqref{eq:STF_dissipative_tensor}, the third (divergence-of-tensor)
contribution on the right-hand side of the general FDDFT
statement~\eqref{eq:fddft_free_energy} vanishes identically via the
orthogonality condition:
\begin{align}\label{eq:dissipative_derivative}
  \int \mathrm d\mathbf{u}' \frac{\delta M(\mathbf{u},\mathbf{u}',t)}{\delta h(\mathbf{u},t)} &= \frac{k_B T}{\eta} \bm{\nabla}_{\mathbf{u}}\cdot h^2(\mathbf{u},t) \int \mathrm d\mathbf{u}' \delta(\mathbf{u}-\mathbf{u}') \bm{\nabla}_{\mathbf{u}'}\delta(\mathbf{u}-\mathbf{u}') \nonumber \\
  &= 0.
\end{align}

From the fluctuation--dissipation
relation~\eqref{eq:fluctuation_dissipation}, the noise amplitude operator is
explicitly evaluated as:
\begin{equation}
    \mathbf{B}(\mathbf{u},\mathbf{u}',t)= \bm{\nabla}_{\mathbf{u}} \cdot \sqrt{\frac{2k_BT h^3(\mathbf{u},t)}{3\eta}} \delta(\mathbf{u}-\mathbf{u}') \mathbf{I}.
\end{equation}
Combining this noise closure with the general FDDFT
statement~\eqref{eq:fddft_free_energy} generates the generalized thin-film
evolution equation:
\begin{align}\label{eq:stf_gradient}
    \partial_t h(\mathbf{u},t)&= \bm{\nabla}_{\mathbf{u}} \cdot \left (\frac{h^3(\mathbf{u},t)}{3\eta} \bm{\nabla}_{\mathbf{u}} \frac{\delta F[h]}{\delta h(\mathbf{u},t)} \right) \nonumber \\
    &+  \bm{\nabla}_{\mathbf{u}} \cdot \left( \sqrt{\frac{2k_BT h^3(\mathbf{u},t)}{3\eta}}\bm{\xi}(\mathbf{u},t)\right).
\end{align}
Defining $Q(h)=h^3/(3\eta)$ as the hydrodynamic mobility,
Eq.~\eqref{eq:stf_gradient} matches the well-known variational gradient
dynamics form~\cite{mitlin1993dewetting,thiele2010thin} augmented by the
appropriate thermal fluctuation term~\cite{duran2019instability}.

The continuous free-energy functional $F[h]$ is expressed in terms of the
surface tension $\gamma$ and the substrate binding potential $\Phi(h)$
as~\cite{Peter2016}:
\begin{equation}\label{eq:thin_film_free_energy}
  F[h]=\int \mathrm d\mathbf{u}\left[\gamma \sqrt{1+|\bm{\nabla}_{\mathbf{u}}h|^2} +\Phi(h) \right],
\end{equation}
where $\sqrt{1+|\bm{\nabla}_{\mathbf{u}}h|^2}\,\mathrm d\mathbf{u}$ defines
the differential surface area element. The binding potential models the
molecular interactions between the substrate and the fluid film, and is
directly related to the disjoining or Derjaguin pressure, $\Pi=-\Phi'(h)$,
valid for parallel or nearly parallel
interfaces~\cite{Frumkin1938,Derjaguin1986,Derjaguin1987}. Physically, $\Pi$
represents the structural excess pressure acting on the solid substrate due
to the presence of the thin liquid
layer~\cite{de1985wetting,Andreas2014,Peter2021}.

In the long-wave limit ($\bm{\nabla}_{\mathbf{u}}h \ll 1$), the area element
simplifies asymptotically to $\sqrt{1+|\bm{\nabla}_{\mathbf{u}}h|^2} \approx
1+ |\bm{\nabla}_{\mathbf{u}}h|^2/2$. Substituting this linearised expansion
into the free-energy functional~\eqref{eq:thin_film_free_energy} and taking
the functional derivative yields the final stochastic thin-film (STF)
equation:
\begin{align}\label{eq:stf}
  \partial_t h(\mathbf{u},t)&= \bm{\nabla}_{\mathbf{u}} \cdot \left (\frac{h^3(\mathbf{u},t)}{3\eta} \bm{\nabla}_{\mathbf{u}} [\Phi'(h)-\gamma \nabla_{\mathbf{u}}^2h(\mathbf{u},t)]  \right) \nonumber \\
  &+ \bm{\nabla}_{\mathbf{u}} \cdot \left(\sqrt{\frac{2k_B T h^3(\mathbf{u},t)}{3\eta}}\bm{\xi}(\mathbf{u},t)\right),
\end{align}
which recovers the structural form established in previous phenomenological
studies~\cite{Davidovitch2005,grun2006thin,duran2019instability}. Crucially,
while existing derivations require a top-down long-wave reduction of the
macroscopic Landau--Lifshitz fluctuating Navier--Stokes
equations~\cite{landau1987fluid,grun2006thin,duran2019instability}, our
approach delivers a rigorous alternative derivation originating directly from
first-principles microscopic dynamics.

\section{Extensions}\label{sec:extensions}
We illustrate how our first-principles FDDFT formalism is naturally amenable
to deriving transport equations for increasingly complex configurations. We
consider two canonical extensions: a multi-layered immiscible film framework
and a single-layer film undergoing non-conserved evaporation or condensation
mass transport.

\subsection{Two-Layer Film}
\begin{figure}[htbp]
  \centering
  \includegraphics[width=0.45\textwidth]{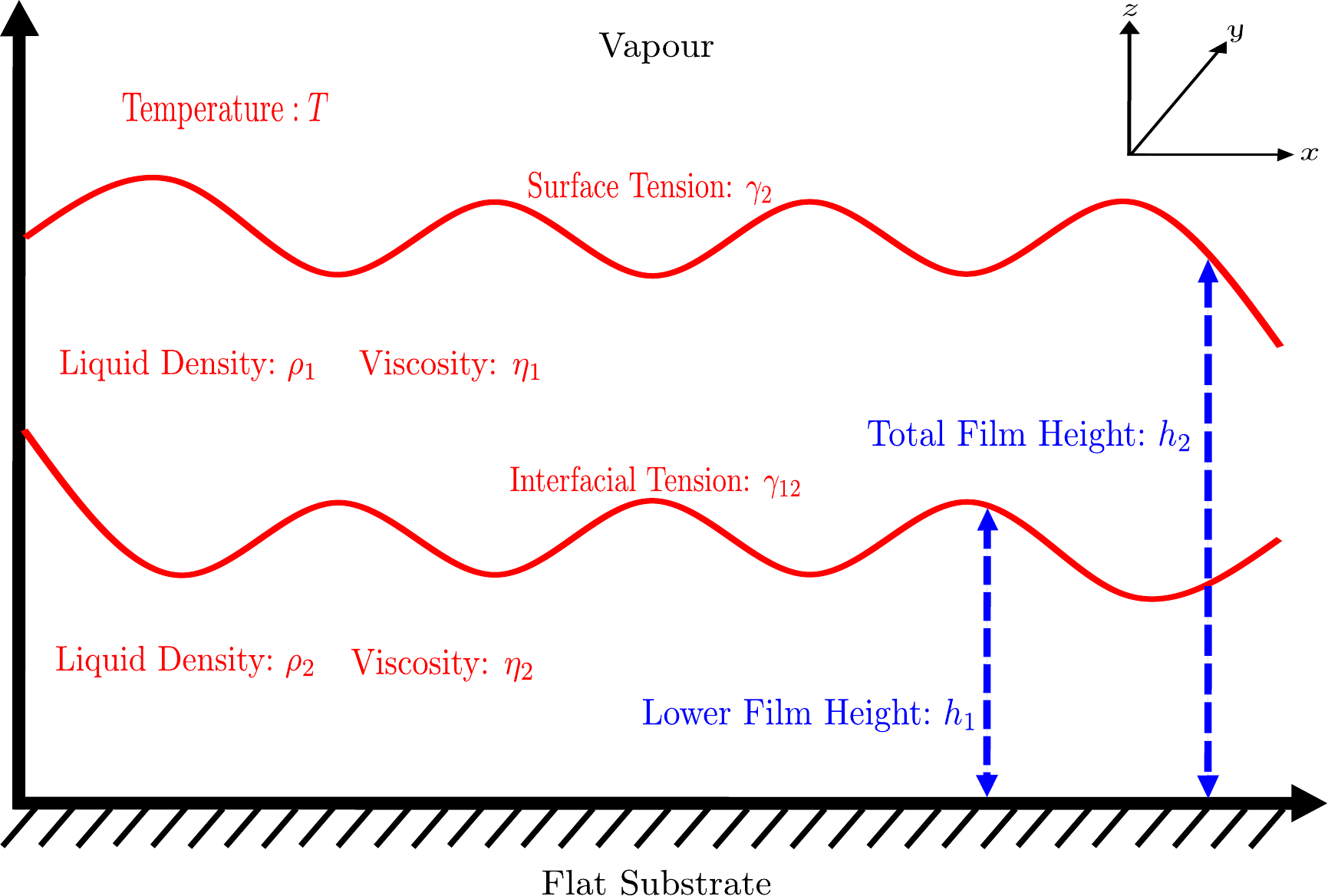}
  \caption{Sketch of the profile geometry of a two-layer film with lower film height $h_1$ and total film height $h_2$. The coordinates $x$ and $y$ map the streamwise and cross-stream directions, with the substrate sitting at the $(x,y)$ plane, and $z$ defining the outward-pointing normal coordinate.}
  \label{fig:two_films}
\end{figure}

Consider a thin fluid layer on a flat substrate matching the Cartesian
coordinate system $(x,y,z)$ at $z=0$, with $x$ and $y$ defining the
streamwise and spanwise coordinates. In this configuration, a second fluid
film of distinct chemical composition lies directly on top of the base layer
in contact with the solid wall (Figure~\ref{fig:two_films}). Assuming the two
fluid phases are strictly immiscible, the lower film interface is located at
$z=h_1(\mathbf{u},t)$, the total combined height from the substrate to the
upper free surface is $z=h_2(\mathbf{u},t)$, and the upper layer thickness is
given by $h_2-h_1$. The resolved mesoscopic variables are mapped by the
vector operator $\hat{\mathbf{Z}}=(\hat{h}_1,\,\hat{h}_2)^\top$. Extending
the single-layer microscopic definition~\eqref{eq:microscopic_height} under
an incompressibility assumption ($\rho_1(\mathbf{r},t)=\rho_1$ for $z \in
[0,h_1)$ and $\rho_2(\mathbf{r},t)=\rho_2$ for $z \in (h_1,h_2]$) yields:
\begin{align}
    \hat{h}_1(\mathbf{u}) &= \frac{1}{\rho_1}\int_0^\infty \hat{\rho}_1(\mathbf{r})\,\mathrm dz,\\
    \hat{h}_2(\mathbf{u}) &= \hat{h}_1(\mathbf{u})+\frac{1}{\rho_2}\int_0^\infty \hat{\rho}_2(\mathbf{r})\,\mathrm dz.
\end{align}

For each observable/fluid phase, we construct the corresponding microscopic
parallel current density:
\begin{equation}
    \hat{\bm{\mathdutchcal{j}}}_j(\mathbf{u})=\frac{1}{\rho_j}\int_0^\infty \mathrm dz \sum_{i=1}^{N_j} (\upsilon_i^{j,x},\,\upsilon_i^{j,y})^\top \delta(\mathbf{r}-\mathbf{q}_i^j),
\end{equation}
which closes the microstate kinematic conservation law via the coupled
sub-relations:
\begin{subequations}
  \begin{align}
  i\mathcal{L}\hat{h}_1(\mathbf{u}) &= -\bm{\nabla}_{\mathbf{u}} \cdot \hat{\bm{\mathdutchcal{j}}}_1(\mathbf{u}), \\
  i\mathcal{L}\hat{h}_2(\mathbf{u})-i\mathcal{L}\hat{h}_1(\mathbf{u}) &= -\bm{\nabla}_{\mathbf{u}} \cdot \hat{\bm{\mathdutchcal{j}}}_2(\mathbf{u}),
\end{align}
\end{subequations}
where the molecular momenta track as $\mathbf{p}_i^1=m_1 \bm{\upsilon}_i^1$
and $\mathbf{p}_i^2=m_2 \bm{\upsilon}_i^2$. Mirroring the single-layer
derivation, the multi-component dissipative tensor blocks expand into the
dual-divergence format
$M_{ij}(\mathbf{u},\mathbf{u}',t)=\nabla_{\mathbf{u}}\cdot
\nabla_{\mathbf{u}'}\cdot \mathbf{D}_{ij}(\mathbf{u},\mathbf{u}',t)$, where
the current-flux Onsager matrices are explicitly defined as:
\begin{subequations}
  \begin{align}
    \mathbf{D}_{11}(\mathbf{u},\mathbf{u}',t) &= \int_0^\tau \mathrm ds \left\langle \hat{\bm{\mathdutchcal{j}}}_1(\mathbf{u}') \otimes \hat{\bm{\mathdutchcal{j}}}_1(\mathbf{u},s)\right\rangle_\text{leq}^t,\\
    \mathbf{D}_{12}(\mathbf{u},\mathbf{u}',t) &= \int_0^\tau \mathrm ds \left\langle (\hat{\bm{\mathdutchcal{j}}}_1(\mathbf{u}')+\hat{\bm{\mathdutchcal{j}}}_2(\mathbf{u}')) \otimes \hat{\bm{\mathdutchcal{j}}}_1(\mathbf{u},s)\right\rangle_\text{leq}^t,\\
    \mathbf{D}_{21}(\mathbf{u},\mathbf{u}',t) &= \int_0^\tau \mathrm ds \left\langle \hat{\bm{\mathdutchcal{j}}}_1(\mathbf{u}') \otimes (\hat{\bm{\mathdutchcal{j}}}_1(\mathbf{u},s)+\hat{\bm{\mathdutchcal{j}}}_2(\mathbf{u},s))\right\rangle_\text{leq}^t,\\
    \mathbf{D}_{22}(\mathbf{u},\mathbf{u}',t) &= \int_0^\tau \mathrm ds \big\langle (\hat{\bm{\mathdutchcal{j}}}_1(\mathbf{u}')+\hat{\bm{\mathdutchcal{j}}}_2(\mathbf{u}')) \nonumber \\
    &\quad \otimes (\hat{\bm{\mathdutchcal{j}}}_1(\mathbf{u},s)+\hat{\bm{\mathdutchcal{j}}}_2(\mathbf{u},s))\big\rangle_\text{leq}^t.
\end{align}
\end{subequations}
Invoking lateral isotropic rotational symmetry and spatial planes locality,
we reduce the multi-layer Onsager matrix kernel to:
\begin{equation}\label{eq:K_full}
    \mathbf{D}_{ij}(\mathbf{u},z,\mathbf{u}',z',t)=A^{-1}\delta(\mathbf{u}-\mathbf{u}')\int_0^\infty \mathrm ds C_{ij}(z,z',t,s) \mathbf{I},
\end{equation}
where the vertically resolved transverse current correlation components
satisfy:
\begin{subequations}
  \begin{align}\label{eq:transverse_components}
    C_{11}(z,z',t,s) &= \frac{1}{2}\int_0^\tau \mathrm ds \left\langle \hat{\bm{\mathdutchcal{j}}}_1(z') \cdot \hat{\bm{\mathdutchcal{j}}}_1(z,s)\right\rangle_\text{leq}^t,\\
    C_{12}(z,z',t,s) &= \frac{1}{2}\int_0^\tau \mathrm ds \left\langle (\hat{\bm{\mathdutchcal{j}}}_1(z')+\hat{\bm{\mathdutchcal{j}}}_2(z')) \cdot \hat{\bm{\mathdutchcal{j}}}_1(z,s)\right\rangle_\text{leq}^t,\\
    C_{21}(z,z',t,s) &= \frac{1}{2}\int_0^\tau \mathrm ds \left\langle \hat{\bm{\mathdutchcal{j}}}_1(z') \cdot (\hat{\bm{\mathdutchcal{j}}}_1(z,s)+\hat{\bm{\mathdutchcal{j}}}_2(z,s))\right\rangle_\text{leq}^t,\\
    C_{22}(z,z',t,s) &= \frac{1}{2}\int_0^\tau \mathrm ds \big\langle (\hat{\bm{\mathdutchcal{j}}}_1(z')+\hat{\bm{\mathdutchcal{j}}}_2(z')) \cdot (\hat{\bm{\mathdutchcal{j}}}_1(z,s)+\hat{\bm{\mathdutchcal{j}}}_2(z,s))\big\rangle_\text{leq}^t.
\end{align}
\end{subequations}

Each transverse component $C_{ij}(z,z',t,s)$ obeys a hydrodynamics diffusion
law:
\begin{equation}
    \partial_s C_{ij}(z,z',t,s) = \nu \partial_z^2 C_{ij}(z,z',t,s),
\end{equation}
where the kinematic viscosity coefficient shifts discontinuously across the
fluid--fluid interface:
\begin{equation}
    \nu=
    \begin{cases}
      \frac{\eta_1}{m_1 \rho_1}, & z\in [0,h_1),\\
      \frac{\eta_2}{m_2 \rho_2}, & z\in (h_1,h_2],
    \end{cases}
\end{equation}
with $\eta_1$ and $\eta_2$ representing the respective phase dynamic
viscosities. Utilising the local canonical equilibrium
distribution~\eqref{eq:gen_canonical_Hamiltonian}, we evaluate the exact
initial state condition boundaries as:
\begin{subequations}\label{eq:two_films_initial}
  \begin{align}
    C_{11}(z,z',t,0) &= \frac{1}{2 \rho_1^2}\left\langle \sum_{ij} \bm{\upsilon}_i^1 \cdot \bm{\upsilon}_j^1 \delta(z-q_i^{1,z}) \delta(z'-q_j^{1,z}) \right\rangle_\text{leq}^t \nonumber \\
    &= \frac{\langle \hat{\rho}_1(z)\rangle_\text{leq}^t}{\beta m_1 \rho_1^2} \delta(z-z') \mathbb{I}_{0<z<h_1},\\
    C_{12}(z,z',t,0) &= C_{21}(z,z',t,0)=C_{11}(z,z',t,0)\label{eq:C12},\\
    C_{22}(z,z',t,0) &= C_{11}(z,z',t,0) +\frac{\langle \hat{\rho}_2(z)\rangle_\text{leq}^t}{\beta m_2 \rho_2^2} \delta(z-z') \mathbb{I}_{h_1<z<h_2},\label{eq:C22}
\end{align}
\end{subequations}
setting the local density weights uniformly to $\langle
\hat{\rho}_i(z)\rangle_\text{leq}^t=\rho_i A$ to simplify the subsequent
integration steps.

The different components of $C_{ij}(z,z',t,s)$ map individual correlations of
distinct microscopic fluid phase fluxes, requiring separate integration
domains. For the base component $C_{11}$, which tracks solely the lower film
current density~\eqref{eq:transverse_components}, all associated fluid
particles are strictly bounded within $z\in[0,h_1)$. The boundary conditions
mirror the single-layer thresholds~\eqref{eq:z0_bc}--\eqref{eq:zh_bc}, namely
$\mathcal{C}_{11}(0,t,s)=0$ and $\partial_z \mathcal{C}_{11}(h_1,t,s)=0$,
where the marginal quantity satisfies:
\begin{equation}
    \mathcal{C}_{11}(z,t,s)=\int_0^\infty \mathrm dz' \frac{m_1 \rho_1 \beta}{A} C_{11}(z,z',t,s),
\end{equation}
and is governed by the localised diffusion equation $\partial_s
\mathcal{C}_{11}(z,t,s) =\nu_1 \partial_z^2 \mathcal{C}_{11}(z,t,s)$.
Replicating the Laplace transform derivation of \S~\ref{sec:stf} yields the
exact base Onsager matrix block:
\begin{equation}
    \mathbf{D}_{11}(\mathbf{u},\mathbf{u}',t)=k_B T \frac{h_1^3(\mathbf{u},t)}{3 \eta_1} \delta(\mathbf{u}-\mathbf{u}') \mathbf{I}.
\end{equation}

Evaluating the cross-component $C_{12}$ requires splitting the integration
across separate fluid spatial subdomains ($z\in[0,h_1)$ and $z\in(h_1,h_2]$)
because the kernel incorporates current correlations spanning both immiscible
phases. We define the piecewise marginal correlations as:
\begin{equation}\label{eq:M12_UL}
    \mathcal{C}_{12}(z,t,s)=
    \begin{cases}
        \mathcal{C}_{12}^L=\int_0^{h_1} \mathrm dz' \frac{m_1 \rho_1 \beta}{A} C_{12}(z,z',t,s), & z \in [0,h_1),\\[5pt]
        \mathcal{C}_{12}^U=\int_{h_1}^{h_2} \mathrm dz' \frac{m_2 \rho_2 \beta}{A} C_{12}(z,z',t,s), & z \in (h_1,h_2],
    \end{cases}
\end{equation}
which are governed by the respective domain diffusion laws:
\begin{subequations}\label{eq:C12_diffusion}
  \begin{align}
    \partial_s \mathcal{C}_{12}^L(z,t,s) &=\nu_1 \partial_z^2 \mathcal{C}_{12}^L(z,t,s),\\
    \partial_s \mathcal{C}_{12}^U(z,t,s) &=\nu_2 \partial_z^2 \mathcal{C}_{12}^U(z,t,s).
  \end{align}
\end{subequations}
From Eq.~\eqref{eq:C12}, the initial conditions are given by
$\mathcal{C}_{12}^L(z,t,0)=1$ and $\mathcal{C}_{12}^U(z,t,0)=0$. We
supplement the system~\eqref{eq:C12_diffusion} with a lower no-slip condition
$\mathcal{C}_{12}^L(0,t,s)=0$ and an upper stress-free condition $\partial_z
\mathcal{C}_{12}^U(h_2,t,s)=0$. At the fluid--fluid interface ($z=h_1$), we
enforce continuous fields and tangential stress matching,
$\mathcal{C}_{12}^L(h_1,t,s)=\mathcal{C}_{12}^U(h_1,t,s)$ and $\nu_1
\partial_z \mathcal{C}_{12}^L(h_1,t,s)= \nu_2 \partial_z
\mathcal{C}_{12}^U(h_1,t,s)$. Applying the Laplace transform to
Eq.~\eqref{eq:C12_diffusion} and evaluating the long-time final-value limits
yields:
\begin{subequations}
  \begin{align}
  \lim_{\tau \to \infty} \int_0^\tau \mathrm ds \mathcal{C}_{12}^L(z,t,s) &=\lim_{\epsilon \to 0} \widetilde{\mathcal{C}}_{12}^L(z,t,\epsilon)=\frac{z (2 h_1-z)}{2 \nu_1},\\
  \lim_{\tau \to \infty} \int_0^\tau \mathrm ds \mathcal{C}_{12}^U(z,t,s) &=\lim_{\epsilon \to 0} \widetilde{\mathcal{C}}_{12}^U(z,t,\epsilon)=\frac{h_1^2}{2 \nu_1}.
  \end{align}
\end{subequations}
Substituting these fields into the spatial projection
definitions~\eqref{eq:K_full} and~\eqref{eq:M12_UL} isolates the
cross-Onsager matrix component via the split multi-line relation:
\begin{align}
    \mathbf{D}_{12}(\mathbf{u},\mathbf{u}',t) &=\frac{k_B T \nu_1}{\eta_1}\delta(\mathbf{u}-\mathbf{u}') \int_0^{h_1}\mathrm dz \int_0^\infty \mathrm ds \mathcal{C}_{12}^L(z,t,s) \mathbf{I} \nonumber \\
    &\quad + \frac{k_B T \nu_2}{\eta_2} \delta(\mathbf{u}-\mathbf{u}') \int_{h_1}^{h_2}\mathrm dz \int_0^\infty \mathrm ds \mathcal{C}_{12}^U(z,t,s) \mathbf{I} \nonumber \\
    &=\frac{k_B T }{ \eta_1} \frac{h_1^2(\mathbf{u},t)}{2}\left(h_2(\mathbf{u},t)-\frac{h_1(\mathbf{u},t)}{3}\right) \delta(\mathbf{u}-\mathbf{u}') \mathbf{I}.
\end{align}
Owing to initial correlation symmetries~\eqref{eq:C12}, the cross-coupling
matrix blocks are symmetric:
\begin{equation}
  \mathbf{D}_{12}(\mathbf{u},\mathbf{u}',t)=\mathbf{D}_{21}(\mathbf{u},\mathbf{u}',t).
\end{equation}

Similarly, calculating $C_{22}$ requires solving the diffusion problem across
separate immiscible subdomains, parameterizing the piecewise marginal
correlations as:
\begin{equation}\label{eq:M22_UL}
    \mathcal{C}_{22}(z,t,s)=
    \begin{cases}
        \mathcal{C}_{22}^L=\int_0^{h_1} \mathrm dz' \frac{m_1 \rho_1 \beta}{A} C_{22}(z,z',t,s), & z \in [0,h_1),\\[5pt]
        \mathcal{C}_{22}^U=\int_{h_1}^{h_2} \mathrm dz' \frac{m_2 \rho_2 \beta}{A} C_{22}(z,z',t,s), & z \in (h_1,h_2],
    \end{cases}
\end{equation}
which obey the coupled domain diffusion laws:
\begin{subequations}\label{eq:C22_diffusion}
  \begin{align}
    \partial_s \mathcal{C}_{22}^L(z,t,s) &=\nu_1 \partial_z^2 \mathcal{C}_{22}^L(z,t,s),\\
    \partial_s \mathcal{C}_{22}^U(z,t,s) &=\nu_2 \partial_z^2 \mathcal{C}_{22}^U(z,t,s).
  \end{align}
\end{subequations}
From Eq.~\eqref{eq:C22}, the uniform initial conditions map as
$\mathcal{C}_{22}^L(z,t,0)=1$ and $\mathcal{C}_{22}^U(z,t,0)=1$. Spanning
both fluid domains, the boundary constraints match the $C_{12}$ configuration
exactly, namely $\mathcal{C}_{22}^L(0,t,s)=0$, $\partial_z
\mathcal{C}_{22}^U(h_2,t,s)=0$, along with the interfacial field and stress
continuities at $z=h_1$:
$\mathcal{C}_{22}^L(h_1,t,s)=\mathcal{C}_{22}^U(h_1,t,s)$ and $\nu_1
\partial_z \mathcal{C}_{22}^L(h_1,t,s)= \nu_2 \partial_z
\mathcal{C}_{22}^U(h_1,t,s)$. Integrating Eq.~\eqref{eq:C22_diffusion} under
these boundary matches yields the relaxed long-time integral profiles:
\begin{subequations}
  \begin{align}
  \lim_{\tau \to \infty} \int_0^\tau \mathrm ds \mathcal{C}_{22}^L(z,t,s) &=\lim_{\epsilon \to 0} \widetilde{\mathcal{C}}_{22}^L(z,t,\epsilon)=\frac{z (2 h_1-z)}{2 \nu_1},\\
  \lim_{\tau \to \infty} \int_0^\tau \mathrm ds \mathcal{C}_{22}^U(z,t,s) &=\lim_{\epsilon \to 0} \widetilde{\mathcal{C}}_{22}^U(z,t,\epsilon) \nonumber \\
  &=\frac{h_1 (h_1-2 h_2) (h_1-h_2)}{2 \nu_1}-\frac{(h_1-h_2)^3}{3 \nu_2}.
  \end{align}
\end{subequations}
Substituting these integrated fields into the projection
operators~\eqref{eq:K_full} and~\eqref{eq:M22_UL} resolves the total Onsager
block:
\begin{align}
    \mathbf{D}_{22}(\mathbf{u},\mathbf{u}',t) &=\frac{k_B T \nu_1}{\eta_1}\delta(\mathbf{u}-\mathbf{u}') \int_0^{h_1}\mathrm dz \int_0^\infty \mathrm ds \mathcal{C}_{22}^L(z,t,s) \mathbf{I} \nonumber \\
    &\quad + \frac{k_B T \nu_2}{\eta_2} \delta(\mathbf{u}-\mathbf{u}') \int_{h_1}^{h_2}\mathrm dz \int_0^\infty \mathrm ds \mathcal{C}_{22}^U(z,t,s) \mathbf{I} \nonumber \\
    &=k_B T \frac{1}{3 \eta_1} \left(\frac{(h_2-h_1)^3(\eta_1-\eta_2)}{\eta_2} +h_2^3\right)\delta(\mathbf{u}-\mathbf{u}') \mathbf{I}.
\end{align}

Mapping this block-Onsager array back onto the generalized dissipative
tensor, the conservative variational structure dictates that the third
(divergence-of-tensor) term in Eq.~\eqref{eq:fddft_free_energy} vanishes
identically. Utilising the fluctuation--dissipation
relation~\eqref{eq:fluctuation_dissipation} over this symmetric tensor
system, we close the coupled stochastic forcing amplitudes $\mathbf{B}_{ij}$.
Inserting these operators directly into the general free-energy-driven FDDFT
framework~\eqref{eq:fddft_free_energy} establishes the stochastic multi-layer
thin-film equations:
\begin{widetext}
\begin{subequations}\label{eq:two_stf}
  \begin{align}
    \partial_t h_1 &= \bm{\nabla}_{\mathbf{u}} \cdot \left[\frac{h_1^3}{3\eta_1} \bm{\nabla}_{\mathbf{u}} \frac{\delta F}{\delta h_1} + \frac{h_1^2}{2 \eta_1 }\left(h_2-\frac{h_1}{3}\right) \bm{\nabla}_{\mathbf{u}} \frac{\delta F}{\delta h_2}\right]+ \bm{\nabla}_{\mathbf{u}} \cdot \left(\sqrt{\frac{2k_B T h_1^3}{3\eta_1}}\bm{\xi}_1(\mathbf{u},t)\right), \label{eq:lower_stf}\\
    \partial_t h_2 &= \bm{\nabla}_{\mathbf{u}} \cdot \left[\frac{h_1^2}{2 \eta_1 }\left(h_2-\frac{h_1}{3}\right) \bm{\nabla}_{\mathbf{u}} \frac{\delta F}{\delta h_1} + \frac{1}{3 \eta_1} \left(\frac{(h_2-h_1)^3(\eta_1-\eta_2)}{\eta_2} +h_2^3\right) \bm{\nabla}_{\mathbf{u}} \frac{\delta F}{\delta h_2}\right] \nonumber \\
    &\quad + \bm{\nabla}_{\mathbf{u}} \cdot \left(\sqrt{\frac{2k_B T h_1}{3\eta_1}}\frac{(3 h_2-h_1)}{2} \bm{\xi}_1(\mathbf{u},t)\right) + \bm{\nabla}_{\mathbf{u}} \cdot \left(\frac{\sqrt{2k_B T}}{2} (h_2-h_1)  \sqrt{\frac{h_1}{\eta_1}+\frac{4 (h_2-h_1)}{3 \eta_2}}\bm{\xi}_2(\mathbf{u},t)\right).\label{eq:total_stf}
  \end{align}
\end{subequations}
\end{widetext}
The collective non-equilibrium free-energy functional $F[h]$ incorporates
both fluid phases, written under the long-wave approximation
as~\cite{pototsky2005morphology,pototsky2006evolution}:
 \begin{equation}\label{eq:two_thin_film_free_energy}
  F[h]=\int \mathrm d\mathbf{u}\left[\gamma_{12} \frac{|\bm{\nabla}_{\mathbf{u}}h_1|^2}{2}+\gamma_{2} \frac{|\bm{\nabla}_{\mathbf{u}}h_2|^2}{2} +\Phi(h_1,h_2) \right],
\end{equation}
where $\gamma_{12}$ is the fluid--fluid interfacial tension, $\gamma_2$ is
the upper film free-surface tension, and $\Phi(h_1,h_2)$ is the multi-layer
interaction potential.

The deterministic mean of Eq.~\eqref{eq:two_stf} recovers the exact classical
long-wave equations derived macroscopically from continuum balance
laws~\cite{danov1998stability,pototsky2004alternative,bandyopadhyay2005instability,pototsky2005morphology,fisher2005nonlinear,pototsky2006evolution,jachalski2013stationary}.
However, as thicknesses scale down to the nanometer regime, thermal noise
introduces crucial stochastic paths. Under the fluctuation-dissipation
relation in~\eqref{eq:fluctuation_dissipation}, the stochastic forcing within
the lower layer $h_1$~\eqref{eq:lower_stf} remains structurally identical to
the single-film instance~\eqref{eq:stf}. Consequently, upper-film
fluctuations affect the lower film exclusively through implicit free-energy
coupling. Conversely, the total height evolution $h_2$~\eqref{eq:total_stf}
features a dual-noise structure: a correlated contribution stemming directly
from lower-layer fluctuations, reflecting the geometric fact that $h_1$ is
embedded within $h_2$, and an independent, emergent noise component
$\bm{\xi}_2(\mathbf{u},t)$ that isolates the localised thermal fluctuations
of the upper phase.

\subsection{Evaporating Film}
We next examine configurations where the film mass is non-conserved, being
modulated by phase-change mechanisms such as evaporation or condensation.
Because the underlying dynamics are non-conserved, the
definition~\eqref{eq:microscopic_height} becomes insufficient due to its
reliance on a negligible vapor phase, which fails when mass transport across
the liquid--vapor interface is explicitly resolved. To remedy this, we
introduce an alternative microscopic height operator that allows for
non-conserved local mass variations:
\begin{equation}\label{eq:microscopic_height_evap}
  \hat{h}(\mathbf{u})=\frac{1}{\rho_l}\int_0^h \mathrm dz \hat{\rho}(\mathbf{r}).
\end{equation}
The integration domain is strictly bounded by the local ensemble-averaged
height $h(\mathbf{u},t)$, ensuring consistency such that $\langle
\hat{h}(\mathbf{u})\rangle_\text{leq}^t=h(\mathbf{u},t)$. Invoking a
timescale separation argument wherein the macroscopic relaxation of
$h(\mathbf{u},t)$ remains slow relative to fast molecular transitions, we
treat the upper limit $h$ as effectively constant during the micro-scaling
integration. This decouples the explicit time-dependence of the operator
$\hat{h}$, permitting the direct deployment of the generalized free-energy
FDDFT framework~\eqref{eq:fddft_free_energy} as established in preceding
sections.

To fully close the non-conserved mass budget, we track a coupled mesoscopic
variable representing the height-averaged vapor density, whose microscopic
definition reads:
\begin{equation}
  \hat{\psi}_v(\mathbf{u})=\int_h^\infty \mathrm dz \hat{\rho}(\mathbf{r}).
\end{equation}
The resolved mesoscopic fields are then gathered in the vector array
$\hat{\mathbf{Z}}=(\hat{h},\,\hat{\psi}_v)^\top$.

Evaluating the time-derivatives of these fields under the Liouvillian
operator yields the microscopic flux conditions. For the height operator, we
state:
\begin{equation}
  i\mathcal{L}\hat{h}(\mathbf{u})=\frac{1}{\rho_l}\int_0^h \mathrm dz i\mathcal{L}\hat{\rho}(\mathbf{r})=- \frac{1}{\rho_l}\int_0^h \mathrm dz \bm{\nabla}_\mathbf{r} \cdot \hat{\mathbf{j}}(\mathbf{r}),
\end{equation}
where $\hat{\mathbf{j}}(\mathbf{r})=\sum_{i=1}^N \bm{\upsilon}_i
\delta(\mathbf{r}-\mathbf{q}_i)$ is the full 3D microscopic velocity flux.
Accounting for the non-uniform spatial variation of the interface boundary
$h(\mathbf{u},t)$, we pull the spatial divergence operator out of the
vertical integral via Leibniz's rule, generating the multi-line relation:
\begin{align}\label{eq:height_evap_derivative}
    i\mathcal{L}\hat{h}(\mathbf{u}) &= - \frac{1}{\rho_l}\bm{\nabla}_{\mathbf{u}}\cdot \hat{\mathbf{j}}^{\parallel}_l(\mathbf{u}) - \frac{1}{\rho_l} \hat{j}^\perp(\mathbf{u},h) +  \frac{1}{\rho_l}(\nabla_{\mathbf{u}} h) \cdot \hat{\mathbf{j}}^{\parallel}(\mathbf{u},h) \nonumber \\
    &= - \frac{1}{\rho_l} \bm{\nabla}_{\mathbf{u}}\cdot \hat{\mathbf{j}}^{\parallel}_l(\mathbf{u}) - \frac{1}{\rho_l} \hat{\mathbf{j}}(\mathbf{u},h) \cdot (-\partial_x h,-\partial_y h,1)^\top \nonumber \\
    &= - \frac{1}{\rho_l} \bm{\nabla}_{\mathbf{u}}\cdot \hat{\mathbf{j}}^{\parallel}_l(\mathbf{u}) - \frac{1}{\rho_l} \hat{\mathbf{j}}(\mathbf{u},h) \cdot \hat{\mathbf{n}}\sqrt{1+|\bm{\nabla}_{\mathbf{u}} h|^2} \nonumber \\
    &= - \frac{1}{\rho_l} \bm{\nabla}_{\mathbf{u}}\cdot \hat{\mathbf{j}}^{\parallel}_l(\mathbf{u}) - \frac{1}{\rho_l} \hat{j}_{\hat{\mathbf{n}}}(\mathbf{u},h)\sqrt{1+|\bm{\nabla}_{\mathbf{u}} h|^2}.
\end{align}
Here, the integrated lateral microscopic momentum flux within the liquid layer is defined as:
\begin{equation}
  \hat{\mathbf{j}}^{\parallel}_l(\mathbf{u})=\int_0^h \mathrm dz \sum_{i=1}^N (\upsilon_i^x,\,\upsilon_i^y)^\top \delta(\mathbf{r}-\mathbf{q}_i).
\end{equation}
The parallel and perpendicular decompositions of the 3D flux vector are denoted by $\hat{\mathbf{j}}^{\parallel}(\mathbf{r})$ and $\hat{j}^\perp(\mathbf{r})$, respectively, such that $\hat{\mathbf{j}}(\mathbf{r})=(\hat{\mathbf{j}}^{\parallel}(\mathbf{r}),\,\hat{j}^\perp(\mathbf{r}))^\top$. The outward-pointing unit normal vector to the curved free surface satisfies:
\begin{equation}
    \hat{\mathbf{n}}=\frac{(-\partial_x h, \, -\partial_y h, \, 1)^\top}{\sqrt{1+|\nabla_{\mathbf{u}} h|^2}},
\end{equation}
isolating the normal interfacial out-flux component via the projection $\hat{j}_{\hat{\mathbf{n}}}(\mathbf{u},h)=\hat{\mathbf{j}}(\mathbf{u},h)\cdot \hat{\mathbf{n}}$.

Applying an identical Leibniz projection to the height-averaged vapor density
operator
\begin{equation}\label{eq:vapor_density_derivative}
    i\mathcal{L}\hat{\psi}_v(\mathbf{u})= -\nabla_{\mathbf{u}}\cdot \hat{\mathbf{j}}^{\parallel}_v(\mathbf{u}) + \hat{j}_{\hat{\mathbf{n}}}(\mathbf{u},h)\sqrt{1+|\nabla_{\mathbf{u}} h|^2},
\end{equation}
resolves the microscopic lateral flux:
\begin{equation}
  \hat{\mathbf{j}}^{\parallel}_v(\mathbf{u})=\int_h^\infty \mathrm dz \sum_{i=1}^N (\upsilon_i^x,\,\upsilon_i^y)^\top \delta(\mathbf{r}-\mathbf{q}_i).
\end{equation}

We evaluate these microscopic relations to construct the dissipative tensor,
decomposing it into its conserved and non-conserved contributions:
$\mathbf{M}(\mathbf{u},\mathbf{u}',t)=\mathbf{M}^c(\mathbf{u},\mathbf{u}',t)+\mathbf{M}^{nc}(\mathbf{u},\mathbf{u}',t)$.
The conservative contribution maps to the parallel current Onsager matrix
format, $M_{ij}^c(\mathbf{u},\mathbf{u}',t)=\nabla_{\mathbf{u}}\cdot
\nabla_{\mathbf{u}'}\cdot \mathbf{D}_{ij}(\mathbf{u},\mathbf{u}',t)$, where
only the first terms on the right-hand sides of
Eqs.~\eqref{eq:height_evap_derivative}
and~\eqref{eq:vapor_density_derivative} remain non-zero. Assuming the
microscopic lateral flux operators in the liquid layer,
$\hat{\mathbf{j}}^{\parallel}_l(\mathbf{u})$, and within the vapor region,
$\hat{\mathbf{j}}^{\parallel}_v(\mathbf{u})$, are uncorrelated, the
cross-coupling Onsager matrices vanish identically,
$\mathbf{D}_{12}=\mathbf{D}_{21}=\bm{0}$. Extending the localised
single-layer result~\eqref{eq:stf_onsager_matrix} directly yields the
liquid-phase block:
\begin{align}\label{eq:K11_evap}
    \mathbf{D}_{11}(\mathbf{u},\mathbf{u}',t) &= \frac{1}{\rho_l^2}\int_0^\tau \mathrm ds \left\langle \hat{\mathbf{j}}^{\parallel}_l(\mathbf{u}') \otimes \hat{\mathbf{j}}^{\parallel}_l(\mathbf{u},s)\right\rangle_\text{leq}^t \nonumber \\
    &= k_B T \frac{h^3(\mathbf{u},t)}{3\eta}\delta(\mathbf{u}-\mathbf{u}')\mathbf{I}.
\end{align}
The vapor-phase block is given by the current-flux correlation integral.
\begin{equation}
  \mathbf{D}_{22}(\mathbf{u},\mathbf{u}',t)=\int_0^\tau \mathrm ds \left\langle \hat{\mathbf{j}}^{\parallel}_v(\mathbf{u}')
  \otimes \hat{\mathbf{j}}^{\parallel}_v(\mathbf{u},s)\right\rangle_\text{leq}^t.
\end{equation}
Rather than recovering the lubrication-scaled mobility profile of
Eq.~\eqref{eq:K11_evap}, the lateral vapor flux is dominated by free
molecular diffusion in the surrounding gas phase. We evaluate the current
block in a wide-text algebraic layout:
\begin{widetext}
  \begin{align}\label{eq:K22_evap}
  \mathbf{D}_{22}(\mathbf{u},\mathbf{u}',t) &= \int_0^\tau \mathrm ds \int_{h}^\infty \mathrm dz' \int_{h}^\infty \mathrm dz \sum_{i,j} \left\langle \bm{\upsilon}_j^{\parallel} \otimes \bm{\upsilon}_i^{\parallel}(s) \delta(\mathbf{r}'-\mathbf{q}_j)\delta(\mathbf{r}-\mathbf{q}_i(s))\right\rangle_{\bm{\lambda_t}} \nonumber \\
  &\approx \int_0^\tau \mathrm ds \int_{h}^\infty \mathrm dz' \int_{h}^\infty \mathrm dz \sum_{i,j} \left\langle \bm{\upsilon}_j^{\parallel} \otimes \bm{\upsilon}_i^{\parallel}(s) \right\rangle_\text{leq}^t \left\langle\delta(\mathbf{r}'-\mathbf{q}_j)\delta(\mathbf{r}-\mathbf{q}_i)\right\rangle_{\bm{\lambda_t}} \nonumber \\
  &\approx \int_{h}^\infty \mathrm dz' \int_{h}^\infty \mathrm dz \sum_{i,j} \mathcal{D}_v \delta_{ij} \mathbf{I} \left\langle\delta(\mathbf{r}'-\mathbf{q}_j)\delta(\mathbf{r}-\mathbf{q}_i)\right\rangle_{\bm{\lambda_t}} = \mathcal{D}_v \psi_v(\mathbf{u}) \delta(\mathbf{u}-\mathbf{u}') \mathbf{I},
\end{align}
\end{widetext}
where $\bm{\upsilon}_i^{\parallel}=(\upsilon_i^x,\,\upsilon_i^y)^\top$. Here,
we assume the particle momentum and position fields are statistically
independent over the integration window, with the positions evolving slowly
enough to be treated as static constraints. We furthermore neglect
cross-particle velocity correlations, corresponding to a mathematically sound
dilute gas limit~\cite{espanol2009derivation,te2020projection}, and close the
lateral diffusion constant $\mathcal{D}_v$ via the single-particle
Green--Kubo formulation under a local equilibrium approximation:
\begin{equation}
  \mathcal{D}_v=\frac{1}{2}\int_0^\tau \mathrm ds \left\langle \bm{\upsilon}_i^{\parallel} \cdot \bm{\upsilon}_i^{\parallel} \right\rangle_\text{eq}.
\end{equation}

To compute the non-conserved contributions to the dissipative tensor, we
isolate the normal interfacial out-flux Green-Kubo autocorrelation from
relations~\eqref{eq:height_evap_derivative}
and~\eqref{eq:vapor_density_derivative}:
\begin{equation}\label{eq:surface_flux_GK}
    \int_0^\tau \mathrm ds \langle \hat{j}_{\hat{\mathbf{n}}}(\mathbf{u}',h) \hat{j}_{\hat{\mathbf{n}}}(\mathbf{u},h,s)\rangle_\text{leq}^t.
\end{equation}
Defining the integrated surface flux as $\hat{j}_{\hat{\mathbf{n}}}(h) =\int
\mathrm d\mathbf{u} \sqrt{1+|\bm{\nabla}_{\mathbf{u}} h|^2}
\hat{j}_{\hat{\mathbf{n}}}(\mathbf{u},h)$, we introduce the macroscopic
evaporative Onsager mobility coefficient $\mathcal{M}_\text{ev}(h,t)$ over
the total interfacial surface area $\mathcal{A}=\int \mathrm
d\mathbf{u}\sqrt{1+|\bm{\nabla}_{\mathbf{u}} h|^2}$
as~\cite{onsager1931reciprocal,onsager1931reciprocal2}:
\begin{equation}
    \mathcal{M}_\text{ev}(h,t)=\frac{1}{k_B T \mathcal{A}} \int_0^\tau \mathrm ds \langle \hat{j}_{\hat{\mathbf{n}}}(h) \hat{j}_{\hat{\mathbf{n}}}(h,s)\rangle_\text{leq}^t.
\end{equation}
Invoking a spatial locality approximation within the tangent planes allows us
to relate Eq.~(\ref{eq:surface_flux_GK}) to the evaporative mobility,
\begin{align}
    &\frac{1}{k_B T} \int_0^\tau \mathrm ds \langle \hat{j}_{\hat{\mathbf{n}}}(h) \hat{j}_{\hat{\mathbf{n}}}(h,s)\rangle_\text{leq}^t = \frac{1}{k_B T} \int \mathrm d\mathbf{u} \int \mathrm d\mathbf{u}' \sqrt{1+|\bm{\nabla}_{\mathbf{u}'} h|^2} \nonumber \\
    &\times \sqrt{1+|\bm{\nabla}_{\mathbf{u}} h|^2} \int_0^\tau \mathrm ds \langle \hat{j}_{\hat{\mathbf{n}}}(\mathbf{u}',h) \hat{j}_{\hat{\mathbf{n}}}(\mathbf{u},h,s)\rangle_\text{leq}^t \nonumber \\
    &\approx \int \mathrm d\mathbf{u} \int \mathrm d\mathbf{u}' \sqrt{1+|\bm{\nabla}_{\mathbf{u}'} h|^2} \mathcal{M}_\text{ev}(h,t) \delta(\mathbf{u}-\mathbf{u}')=\mathcal{M}_\text{ev}(h,t) \mathcal{A},
\end{align}
yielding the localised correlation signature:
\begin{equation}\label{eq:GK_mobility}
    \int_0^\tau \mathrm ds \langle \hat{j}_{\hat{\mathbf{n}}}(\mathbf{u}',h) \hat{j}_{\hat{\mathbf{n}}}(\mathbf{u},h,s)\rangle_\text{leq}^t= k_B T \frac{\mathcal{M}_\text{ev}(h,t)}{\sqrt{1+|\bm{\nabla}_{\mathbf{u}} h|^2}} \delta(\mathbf{u}-\mathbf{u}').
\end{equation}
Combining
fields~\eqref{eq:height_evap_derivative},~\eqref{eq:vapor_density_derivative},
and~\eqref{eq:GK_mobility} resolves the non-conservative dissipative tensor
array blocks:
\begin{equation}\label{eq:dissipative_non_conservative}
    \mathbf{M}^{nc}(\mathbf{u},\mathbf{u}',t)=k_B T \sqrt{1+|\bm{\nabla}_{\mathbf{u}} h|^2} \mathcal{M}_\text{ev}(h,t) \left(
    \begin{matrix}
        \frac{1}{\rho_l^2} & \frac{-1}{\rho_l}\\
        \frac{-1}{\rho_l} & 1
    \end{matrix}
    \right)\delta(\mathbf{u}-\mathbf{u}').
\end{equation}

Having constructed the dissipative tensor, we evaluate the third term on the
right-hand side of the generalized FDDFT
relation~\eqref{eq:fddft_free_energy}. As established in the single-layer
case~\eqref{eq:dissipative_derivative}, the functional derivative of the
conservative part vanishes identically, $\delta
\mathbf{D}_{ij}^c(\mathbf{u},\mathbf{u}',t)/\delta
\mathbf{z}_j(\mathbf{u}',t)=0$. For the non-conservative part, because the
macroscopic film height field $h$ relaxes slowly relative to the fast
molecular transit fluxes, we treat the evaporative mobility as locally
constant, $\partial_h \mathcal{M}_\text{ev}(h,t)=0$. Evaluating the
functional derivative of $\mathbf{D}^{nc}$ thus reduces strictly to:
\begin{equation}
  \frac{\delta }{\delta h(\mathbf{u}',t)}\sqrt{1+|\bm{\nabla}_{\mathbf{u}} h|^2}=- \bm{\nabla}_{\mathbf{u}'} \cdot \frac{\bm{\nabla}_{\mathbf{u}'} h}{\sqrt{1+|\bm{\nabla}_{\mathbf{u}'} h|^2}}\delta(\mathbf{u}-\mathbf{u}'),
\end{equation}
which vanishes identically upon spatial integration, yielding no net
contribution from the divergence-of-tensor term to the final transport
equations.

Substituting the resolved tensor
components~\eqref{eq:K11_evap},~\eqref{eq:K22_evap},
and~\eqref{eq:dissipative_non_conservative} into the general free-energy
FDDFT framework~\eqref{eq:fddft_free_energy} provides the coupled transport
equations:
\begin{subequations}\label{eq:evap_stf_gen_noise}
  \begin{align}
    \partial_t h &= \bm{\nabla}_{\mathbf{u}} \cdot \left(\frac{h^3}{3\eta} \bm{\nabla}_{\mathbf{u}} \frac{\delta F}{\delta h}\right) \nonumber \\
    &\quad - \sqrt{1+|\bm{\nabla}_{\mathbf{u}} h|^2}\frac{\mathcal{M}_\text{ev}(h,t)}{\rho_l}\left(\frac{1}{\rho_l} \frac{\delta F}{\delta h}-\frac{\delta F}{\delta \psi_v}\right) +\theta_1(\mathbf{u},t),\\
    \partial_t \psi_v &= \bm{\nabla}_{\mathbf{u}} \cdot \left(\tilde{\mathcal{D}}_v \psi_v \bm{\nabla}_{\mathbf{u}} \frac{\delta F}{\delta \psi_v}\right) \nonumber \\
    &\quad + \sqrt{1+|\bm{\nabla}_{\mathbf{u}} h|^2}\mathcal{M}_\text{ev}(h,t)\left(\frac{1}{\rho_l} \frac{\delta F}{\delta h}-\frac{\delta F}{\delta \psi_v}\right) + \theta_2(\mathbf{u},t),
\end{align}
\end{subequations}
where $\tilde{\mathcal{D}}_v = \beta \mathcal{D}_v$. The combined stochastic
forcing terms are defined as $\theta_i(\mathbf{u},t)=\sum_j \int \mathrm
d\mathbf{r}' \mathbf{B}_{ij}(\mathbf{r},\mathbf{r}',t) \cdot
\bm{\xi}_j(\mathbf{r}',t)$, which satisfy the fluctuation--dissipation
correlation structure in a wide-text layout:
\begin{widetext}
  \begin{subequations}\label{eq:theta_correlation}
  \begin{align}
    \mathbb{E}\left[\theta_1(\mathbf{u},t) \theta_1(\mathbf{u}',t') \right] &= 2k_B T\delta(t-t')\left[\bm{\nabla}_{\mathbf{u}} \cdot \left(\frac{h^3(\mathbf{u},t)}{3\eta} \bm{\nabla}_{\mathbf{u}'} \delta(\mathbf{u}-\mathbf{u}')\right) + \sqrt{1+|\bm{\nabla}_{\mathbf{u}} h|^2}\frac{\mathcal{M}_\text{ev}(h,t)}{\rho_l^2}\delta(\mathbf{u}-\mathbf{u}')\right],\\
    \mathbb{E}\left[\theta_2(\mathbf{u},t) \theta_2(\mathbf{u}',t') \right] &= 2k_B T\delta(t-t')\left[\bm{\nabla}_{\mathbf{u}} \cdot \left(\tilde{\mathcal{D}}_v\psi_v(\mathbf{u},t) \bm{\nabla}_{\mathbf{u}'} \delta(\mathbf{u}-\mathbf{u}')\right) + \sqrt{1+|\bm{\nabla}_{\mathbf{u}} h|^2}\mathcal{M}_\text{ev}(h,t)\delta(\mathbf{u}-\mathbf{u}')\right],\\
    \mathbb{E}\left[\theta_1(\mathbf{u},t) \theta_2(\mathbf{u}',t') \right] &= \mathbb{E}\left[\theta_2(\mathbf{u},t) \theta_1(\mathbf{u}',t') \right]=-2k_B T \rho_l^{-1}\sqrt{1+|\bm{\nabla}_{\mathbf{u}} h|^2}\mathcal{M}_\text{ev}(h,t)\delta(\mathbf{u}-\mathbf{u}')\delta(t-t').
  \end{align}
  \end{subequations}
\end{widetext}

Invoking the stochastic principle that complex noise terms can be exactly
represented by an equivalent set of white noise components possessing
identical spectral autocorrelations~\cite{gardiner1985handbook}, we recast
the system~\eqref{eq:evap_stf_gen_noise} in its definitive, explicitly
resolved form:
\begin{widetext}
  \begin{subequations}\label{eq:evap_stf}
      \begin{align}
        \partial_t h &= \bm{\nabla}_{\mathbf{u}} \cdot \left(\frac{h^3}{3\eta} \bm{\nabla}_{\mathbf{u}} \frac{\delta F}{\delta h}\right) - \sqrt{1+|\bm{\nabla}_{\mathbf{u}} h|^2}\frac{\mathcal{M}_\text{ev}(h,t)}{\rho_l}\left(\frac{1}{\rho_l} \frac{\delta F}{\delta h}-\frac{\delta F}{\delta \psi_v}\right) + \bm{\nabla}_{\mathbf{u}} \cdot \left(\sqrt{2k_B T h^3/3\eta}\bm{\xi}_1(\mathbf{u},t)\right) \nonumber \\
        &\quad + \rho_l^{-1} \sqrt{2k_B T \mathcal{M}_\text{ev}(h,t)} \sqrt{1+|\bm{\nabla}_{\mathbf{u}} h|^2} \zeta_h(\mathbf{u},t), \label{eq:evap_stf_h} \\[5pt]
        \partial_t \psi_v &= \bm{\nabla}_{\mathbf{u}} \cdot \left(\tilde{\mathcal{D}}_v \psi_v \bm{\nabla}_{\mathbf{u}} \frac{\delta F}{\delta \psi_v}\right) + \sqrt{1+|\bm{\nabla}_{\mathbf{u}} h|^2}\mathcal{M}_\text{ev}(h,t)\left(\frac{1}{\rho_l} \frac{\delta F}{\delta h}-\frac{\delta F}{\delta \psi_v}\right) + \bm{\nabla}_{\mathbf{u}} \cdot \left(\sqrt{2k_B T \tilde{\mathcal{D}}_v \psi_v}\bm{\xi}_2(\mathbf{u},t)\right) \nonumber \\
        &\quad -  \sqrt{2k_B T \mathcal{M}_\text{ev}(h,t)} \sqrt{1+|\bm{\nabla}_{\mathbf{u}} h|^2} \zeta_h(\mathbf{u},t).\label{eq:evap_stf_psi}
      \end{align}
  \end{subequations}
\end{widetext}

In Eq.~\eqref{eq:evap_stf}, the stochastic forcing is systematically
decomposed into two distinct physical mechanisms. The standard vector fields
$\bm{\xi}_i(\mathbf{u},t)$ correspond strictly to conservative, bulk-flow
thermal fluctuations and remain uncorrelated. Conversely, the scalar noise
field $\zeta_h(\mathbf{u},t)$ maps the emergent fluctuations driven
exclusively by the non-conserved phase-change transport of evaporation and
condensation. This non-conserved noise enters the coupled evolution
equations~\eqref{eq:evap_stf_h} and~\eqref{eq:evap_stf_psi} with opposing
signs, directly satisfying microscopic mass conservation across the phase
boundary. The scalar field defines a surface-projected white noise satisfying
the exact geometric correlation:
\begin{equation}
    \mathbb{E}\left[ \zeta_h(\mathbf{u},t)  \zeta_h(\mathbf{u}',t') \right]=\frac{\delta(\mathbf{u}-\mathbf{u}')}{\sqrt{1+|\bm{\nabla}_{\mathbf{u}} h|^2}}\delta(t-t'),
\end{equation}
demonstrating that the stochastic action $\sqrt{1+|\bm{\nabla}_{\mathbf{u}}
h|^2} \zeta_h(\mathbf{u},t)$ represents a true surface-bound white noise
localised to the liquid--vapor interface surface -- with
$\zeta_h(\mathbf{u},t)$ defined on the curved film surface $h(\mathbf{u},t)$
and projected onto a flat Euclidean surface -- and tied intrinsically to the
non-equilibrium evaporative mobility: $\sqrt{1+|\bm{\nabla}_{\mathbf{u}}
h|^2} \zeta_h(\mathbf{u},t)$ can be interpreted as ``surface white noise''
implying that fluctuations are confined to the liquid-vapor interface of the
film, and are directly related to the mobility of evaporation.

The deterministic mean of Eq.~\eqref{eq:evap_stf} matches the transport
framework recently developed in Ref.~\onlinecite{hartmann2023sessile}, which
explicitly features the metric curvature factor
$\sqrt{1+|\bm{\nabla}_{\mathbf{u}} h|^2}$ modulating the non-conserved
phase-change velocity. Near thermodynamic equilibrium, we treat the
evaporative mobility as a localised constant ($\mathcal{M}_\text{ev}(h,t)
\approx \mathcal{M}_\text{ev}$) and approximate the functional derivative of
the vapor free energy by a uniform reservoir chemical potential, $\delta
F/\delta \psi_v = \mu_v$. Enforcing the long-wave limit
($\bm{\nabla}_{\mathbf{u}} h \ll 1$) simplifies the geometric element
asymptotically to $\sqrt{1+|\bm{\nabla}_{\mathbf{u}} h|^2} \approx 1$. This
decoupling isolates the thin-film height from the background vapor field,
yielding the classic stochastic thin-film equation with non-conserved
evaporative dynamics:
\begin{align}\label{eq:evap_stf_equilibrium}
    \partial_t h &= \bm{\nabla}_{\mathbf{u}} \cdot \left(\frac{h^3}{3\eta} \bm{\nabla}_{\mathbf{u}} \frac{\delta F[h]}{\delta h}\right) - \frac{\mathcal{M}_\text{ev}}{\rho_l}\left(\frac{1}{\rho_l} \frac{\delta F[h]}{\delta h(\mathbf{u},t)}-\mu_v\right) \nonumber \\
    &\quad + \bm{\nabla}_{\mathbf{u}} \cdot \left(\sqrt{\frac{2k_B T h^3}{3\eta}}\bm{\xi}_1(\mathbf{u},t)\right) + \rho_l^{-1} \sqrt{2k_B T \mathcal{M}_\text{ev}}  \zeta(\mathbf{u},t),
\end{align}
where the underlying free-energy functional $F[h]$ is structurally closed by
Eq.~\eqref{eq:thin_film_free_energy}. In this linearised configuration, the
surface-projected thermal forcing operator converges to standard scalar
space-time white noise satisfying the exact flat Euclidean correlation
$\mathbb{E}[\zeta(\mathbf{u},t)  \zeta(\mathbf{u}',t')
]=\delta(\mathbf{u}-\mathbf{u}')\delta(t-t')$. The deterministic portion of
Eq.~\eqref{eq:evap_stf_equilibrium} recovers the familiar lubrication models
widely deployed across thin-film literature (e.g.,
Refs.~\onlinecite{thiele2010thin,thiele2011note,fravstia2011dynamical,fravstia2012modelling,thiele2018recent,hartmann2025gradient}).

\section{Concluding remarks}\label{sec:conclusion}
We have developed a systematic, first-principles derivation of the stochastic
thin-film equation from exact underlying microscopic Hamiltonian dynamics by
leveraging a generalized FDDFT framework. This bottom-up strategy maps the
macroscopic interface evolution onto a designated set of slow microscopic
observables, closed here via a circular definition of the local film height
field. By explicitly evaluating the Green--Kubo dissipative tensor alongside
the fluctuation--dissipation relation and a localised free-energy functional,
we rigorously recover the structural form of the stochastic thin-film
equation. Crucially, while existing models originate from phenomenological,
top-down fluctuating Navier--Stokes descriptions, our formulation establishes
a direct micro-to-meso scaling linkage. Furthermore, this framework offers
exceptional mathematical flexibility, which we demonstrate by seamlessly
deriving novel stochastic thin-film equations for increasingly complex
scenarios, including multi-layered immiscible interfaces and non-conserved
mass transport driven by evaporation and condensation.

Future endeavors will focus on extending this predictive framework to higher
levels of multi-scale physical complexity. Immediate pathways include
investigating thin-film dynamics in the presence of classical and
non-classical nucleation by exploiting previously established FDDFT
nucleation formalisms~\cite{lutsko2012dynamical,duran2018general}. Similarly,
the framework is natural for modeling complex multi-component fluid systems,
including colloidal suspensions~\cite{thiele2011note,thiele2013gradient},
polymeric binary mixtures~\cite{clarke2005toward}, and surfactant-laden
interfaces~\cite{thiele2014patterned,thiele2016gradient}. For
surfactant-driven flows, resolving the emergent Marangoni forces breaks
isothermal constraints, necessitating an entropy rather than a free-energy
functional. Because the fundamental FDDFT statement~\eqref{eq:fddft_entropy}
is formulated strictly in terms of a structural entropy functional, it offers
a mathematically sound, unified vehicle to tackle these non-isothermal open
challenges in subsequent studies.

\begin{acknowledgments}
We acknowledge financial support by the ERC-EPSRC Frontier Research Guarantee
through Grant No. EP/X038645, ERC through Advanced Grant No. 247031, and
EPSRC through Grant Nos. EP/L025159 and EP/L020564.

\end{acknowledgments}

\section*{Data Availability Statement}

Data sharing is not applicable to this article as no new data were created or
analysed in this study.

\bibliography{References}

\end{document}